\documentclass[aps,a4paper,showpacs,preprintnumbers,floatfix,twocolumn,prd]{revtex4-2}

\usepackage{epsfig}
\usepackage{color}
\usepackage{float}
\usepackage{placeins}
\usepackage{appendix}
\usepackage{placeins}
\usepackage{url} 
\usepackage[hidelinks]{hyperref}
\usepackage{comment}

\usepackage[utf8]{inputenc}

\def\markthis{}

\definecolor{red}{rgb}{1.0,0.0,0.0}\def\red{}
\definecolor{blue}{rgb}{0.0,0.0,1.0}

\begin{document}

%

\title{{\red Spontaneous damage annealing reactions as a possible source} of low energy excess in semiconductor detectors}

\author{Kai Nordlund}
\email{kai.nordlund@helsinki.fi}
\affiliation{Department of Physics, University of Helsinki, 
                      P.O.Box 64, FI-00014 University of Helsinki, Finland}
\affiliation{Helsinki Institute of Physics, 
                      P.O.Box 64, FI-00014 University of Helsinki, Finland}

\author{Fanhao Kong}
\email{fanhao.kong@helsinki.fi}
\affiliation{Department of Physics, University of Helsinki, 
                      P.O.Box 64, FI-00014 University of Helsinki, Finland}
\affiliation{Helsinki Institute of Physics, 
                      P.O.Box 64, FI-00014 University of Helsinki, Finland}

\author{Flyura Djurabekova}
\email{flyura.djurabekova@helsinki.fi}
\affiliation{Department of Physics, University of Helsinki, 
                      P.O.Box 64, FI-00014 University of Helsinki, Finland}
\affiliation{Helsinki Institute of Physics, 
                      P.O.Box 64, FI-00014 University of Helsinki, Finland}
                      
\author{Matti Heikinheimo}
\email{matti.heikinheimo@helsinki.fi}
\affiliation{Department of Physics, University of Helsinki, 
                      P.O.Box 64, FI-00014 University of Helsinki, Finland}
\affiliation{Helsinki Institute of Physics, 
                      P.O.Box 64, FI-00014 University of Helsinki, Finland}

\author{Kimmo Tuominen}
\email{kimmo.i.tuominen@helsinki.fi}
\affiliation{Department of Physics, University of Helsinki, 
                      P.O.Box 64, FI-00014 University of Helsinki, Finland}-
\affiliation{Helsinki Institute of Physics, 
                      P.O.Box 64, FI-00014 University of Helsinki, Finland}

\author{Nader Mirabolfathi}                      
\email{mirabolfathi@physics.tamu.edu} 
\affiliation{Department of Physics and Astronomy and Mitchell Institute for Fundamental Physics and Astronomy,
Texas A\&M University, College Station, TX 77843, USA}

\author{Antti Kuronen}
\email{antti.kuronen@helsinki.fi}
\affiliation{Department of Physics, University of Helsinki, 
                      P.O.Box 64, FI-00014 University of Helsinki, Finland}
\affiliation{Helsinki Institute of Physics, 
                      P.O.Box 64, FI-00014 University of Helsinki, Finland}

\date{\today}


\preprint{Under revision in Physical Review Materials (2025)}


\begin{abstract}
{
{\red In semiconductor detectors designed for capturing dark matter particles or neutrinos, when the detection threshold is constantly improved to increasingly low energies,} an ``excess" signal of apparent energy release events below a few hundred eV is observed in several different kinds of detectors.  { \red This becomes a big obstacle to the observation of actual dark matter signals, hindering the detectors' sensitivity for rare events in this energy range.} Using atomistic simulations with a classical thermostat and a quantum thermal bath,
we show that this kind of signal is consistent with energy release from long-term {\red annealing} events  of complex defects that can be formed by any kind of nuclear recoil radiation events. Such {\red energy releases are}  shown to have a very similar exponential dependence {\red on energy release magnitudes} as that observed in experiments.
By detailed analysis of the annealing events, we show that {\red crossing very low energy barriers} can trigger larger energy releases in an avalanche-like effect. This explains why large energy release events can occur even down to cryogenic temperatures, {\red where the significant migration of point defects in silicon is hardly ever possible.}
}
\end{abstract}

\maketitle

\section{Introduction}
\label{sec:intro}

One possible approach to detect light particle dark matter is by looking for low-energy recoil events in materials 
that cannot be attributed to conventional matter radiation events \cite{Arcadi:2017kky, Roszkowski:2017nbc,Kad17}. Similar methods are also used in coherent neutrino nucleus scattering experiments~\cite{Adamski:2024yqt,NUCLEUS:2019igx}. {\red This strong demand from physics research has motivated the consistent development of new detector technologies with lower and lower detection threshold.}

When the detection threshold in semiconductor detectors has been pushed to increasingly low energies, an ``excess'' signal of apparent energy release events below a few hundred eV  is observed in several different kinds of detectors \cite{CRESST:2019axx,CRESST:2019jnq,DAMIC:2020cut,EDELWEISS:2019vjv,EDELWEISS:2020fxc,CRESST:2017ues,NUCLEUS:2019kxv,SENSEI:2020dpa,SuperCDMS:2018mne,SuperCDMS:2020ymb,Fuss:2022fxe,Angloher:2022pas}, {\red being able to mimic or cover a possible DM signal. Sharing a similar energy range with possible dark matter recoil signals, this excess will mask any potential dark matter signal, and therefore understanding its origin and finding out possible ways to mitigate or even eliminate it are questions of utmost importance.} Several different explanations have been offered to this excess, for instance electronic noise \cite{Abbamonte:2022rfh}, stress effects/crack relaxation \cite{Anthony-Petersen:2022ujw,Romani:2024qgw} or solid state effects \cite{Chang:2025gkn}. However, some recent
experiments disfavor an explanation due to {\red intrinsic effects in the crystal or external radiation} \cite{Angloher:2022pas}. It seems {\red somewhat convincing }that there are several components and several different causes of the excess events.

  Although there are thus many uncertainties regarding the origin of the effect,
on a general level we note that the excess signal is observed in detectors consisting of different materials, manufacturing techniques and different electronics with a very similar exponential energy dependence $\propto \exp(-\alpha E)$, where $\alpha>0$ is a constant. This outward phenomenon suggests that there may be an inherent mechanism for the observed excess signal, {\red due to this very obvious similarity in the thermodynamic patterns of the energy release spectra in distinct experiments. Hypothesizing the existence of a common origin is also consistent with the evidence pointing to a solid-state effect \cite{Chang:2025gkn} as the source of the signal. This inherent mechanism should not depend on any specific choices in detector materials, the specific fabrication techniques of detector materials, or the design of the electric circuits in the detectors, and cannot be weakened by the modifications in these very aspects in the detectors. This character of invariance really attracts researchers to dig deeper for the identification of the actual source behind the scene. Even if the investigations are still ongoing and have not reached any consensus on this, the academic gains along this discovery path have already been quite fruitful \cite{CRESST:2019axx,CRESST:2019jnq,DAMIC:2020cut,EDELWEISS:2019vjv,EDELWEISS:2020fxc,CRESST:2017ues,NUCLEUS:2019kxv,SENSEI:2020dpa,SuperCDMS:2018mne,SuperCDMS:2020ymb,Fuss:2022fxe,Angloher:2022pas}, having very profound influence in dark matter detection experiments and science or engineering technology as a whole.}

In this article, we analyze the possibility that the excess signal could originate from {\red local annealing} of nanometer-size range defect clusters. It is well established from both experiments \cite{Rua84,Jen95,Kyu99,Par00} and simulations \cite{Cat96,Nor97f,Hen97,San07c,Hol08a} that keV recoils in semiconductor materials such as Si produce small disordered zones in the material.  Even if the material is not irradiated explicitly, cosmic rays or decay of radioactive impurities that are always present at some concentration in a material and its immediate surrounding environment will produce similar damage pockets. Even if the initial kinetic energy of a radioactive decay event may be in the MeV range, the final damage will be produced by $\sim$ keV recoils, since collision cascades at much higher energies split up into sub-cascades \cite{Ave98,Nor97f,Bac16}.

The nanometric damage regimes produced in such events are higher in potential energy than the pristine crystal structure, {\red 
because of the existence of overcoordination of atoms, undercoordination of atoms and all kinds of deviations from the tetrahedral structures in the disordered damaged region.} Hence, these damages are expected to thermally {\red anneal} towards the ground state at a time scale depending on temperature.  {\red The annealing, which may occur stepwise}, will thus be associated with {\red energy release events.} {\red We note that in the traditional theory of radiation damage, the damage is assumed to be distributed in well-defined isolated interstitials and vacancies, which have a well-defined formation energy and hence if they meet and recombine, the energy release would be always well-defined.
However, numerous studies show that the damage in covalently bonded semiconductors is much more complex 
\cite{Cat96,Nor97f,Mar05} and hence there may be a wide range of different energy release values. For this reason, we tend to avoid using the word ``recombination'' to describe the energy release events.
} 
While there have been previous works analyzing the {\red annealing} of radiation-induced disordered regions \cite{Cat96,Hen97,Web99,Bel13,Bel15,Jay17}, 
 {\red these have not systematically analyzed the energy magnitude distribution of the potential energy release events in annealing. Also, very, very few of them, if any, looked into what happened at cryogenic temperatures, as this is not a relevant temperature regime in industrial applications like semiconductor chip manufacturing.} In this article, we analyze this energy release magnitudes systematically, and show that it leads to {\red energy release} events with an exponential energy dependence, very similar to that found in experiments.

\section{Methods}
\label{sec:methods}

To examine the production of disordered damage zones and their possible energy release on extended time in annealing, we used a five-stage simulation and analysis procedure: (i) simulation of typical radiation damage creation, (ii) simulation of 
thermal annealing of this damage on long time scales, (iii) quenching of configurations to 0 K to obtain potential energy release without effects from kinetic energy fluctuation, (iv) obtaining the energy release in {\red annealing} events from the time dependence of the potential energy in stage (iii), and (v) analysis of the time scale of annealing.
These results are presented in section \ref{sec:results}.
Since at cryogenic temperatures quantum mechanical zero point vibrations can affect atom dynamics, we also employed the quantum thermal bath (QTB) \cite{dammak2009quantum, Bar11} to assess how quantum effects affect the annealing.
The basic ideas of utilizing QTB are introduced in section \ref{subsec:quantummd}, and the results of these simulations are included along with the fully
classical ones in the following discussions.
Finally, to understand why annealing is observed even at cryogenic temperatures, we also analyzed what kind of atomic movements in individual {\red annealing events (energy release events) actually }lead to the energy release (section \ref{sec:annmech} and section \ref{sec:location_energy_release}).

\subsection{Simulation and analysis stages}
\label{subsec:classicalmd}

In all simulation stages (i-v), we used classical molecular dynamics \cite{Allen-Tildesley}.
{\red As for classical molecular dynamics simulations, several of them were carried out with the code PARCAS \cite{PARCAS}.} This code has been extensively used to examine radiation effects
in a wide range of materials, and the results have been shown to agree well with experiments
in many different measurable quantities \cite{Nor97f,Nor98,Nor05c,Hol10a,Norr10,Sch20}. {\red In comparison to other popular molecular dynamics simulators (for example, LAMMPS), PARCAS has a unique advantage of possessing quite a number of choices for accurate repulsive interatomic potentials, which are important when doing careful research on irradiation damage creation by irradiation cascades. 
Besides the classical simulations conducted by PARCAS, a number of other classical cases were modeled with the widely
used LAMMPS molecular dynamics code \cite{LAMMPS} and gave the same
behaviour as the PARCAS results.} The simulations with the quantum thermal bath (QTB) were all done with the LAMMPS codes, {\red since PARCAS does not include this QTB feature.} Details on the specific quantum aspects of the modelling are given in subsection \ref{subsec:quantummd}.
The analysis of annealing mechanisms in section \ref{sec:annmech} were done with the LAMMPS
code since this code has a built-in climbing image nudged elastic band (CI-NEB) \cite{NEB} feature that could be used in analyzing the energy barriers, {\red which is an essential part in the evaluation of kinetics of annealing events.}

The interatomic interactions {\red between silicon atoms} were described with the Stillinger-Weber potential (SW potential) \cite{Sti85}, that had been found to describe well a wide range of properties in Si when compared to experiments \cite{Sam06,Ham19}, DFT calculations \cite{Hol08a} and results from modern machine-learned potentials \cite{Ham19}.  We also tested the Tersoff III potential \cite{Ter88c}, which gave very similar results regarding the energy release behavior. {\red The Tersoff III potential was used as a supplement in our research due to a few reasons. First, the development of this and the SW potential potential were two completely independent processes, so Tersoff results can serve as a comparison with SW results. Second, Tersoff III potential has been used quite intensively in irradiation cascade simulations for silicon \cite{Nor97f}. Moreover, in the PARCAS code, this potential has been augmented by a DMol repulsive potential \cite{Nor96c,Nor25} using the Fermi function joining formalism \cite{Nor96b}.

In the cascade runs done by PARCAS code, the Stillinger-Weber potential was also augmented by the DMol repulsive potential \cite{Nor96c,Nor25}. The cascade runs carried out with the LAMMPS code did not use the repulsive potential, since the code does not support adding a repulsive part to the Stillinger-Weber potential. However, we ran some of the LAMMPS annealing runs based on the PARCAS cascade results, using the output geometries from PARCAS cascade siulations as the input for annealing simulations conducted by LAMMPS. The results on the energy release behavior were in nature identical with pure LAMMPS runs within the statistical error bars.}
In the following parts of the paper, unless otherwise mentioned, the results shown are obtained with the Stillinger-Weber potential, and it is our main tool in this investigation.

In stage (i), we simulated the production of damage by recoil events in Si by simulating 5 keV cascades induced by self-recoils in bulk silicon.  
The energy of 5 keV is above the energy at which cascades break into sub-cascade in Si, i.e. the specific damage pockets produced by higher energy recoils are similar to those produced by 5 keV recoils \cite{Nor97f,Bac18}. The cascades were modelled following standard approaches \cite{Nor97f,Nor18}. 
Briefly, the cascades were initiated in random directions in 3 dimensions  (polar angle $\theta$ and azimuthal angle $\phi$)  with equal weighting of different directions in terms of the solid angle and the development of the cascade was simulated until the system was cooled down back to the ambient temperature, {\red namely, 0 K in the cascade phase.} No temperature or pressure scaling was used in the collisional regions, and heat was removed at the system boundaries, using a Berendsen thermostat applied to a layer of atoms on the boundaries.  {\red Importantly, to properly describe high-energy interactions, we merged the DFT-based ``DMol'' potential at short interatomic separations \cite{Nor96c,Nor25} using the Fermi function smooth joining formalism \cite{Nor96}.} {\red For different temperature levels in annealing (described in the following parts), the number of recoil cases considered ranged between 10 and 1000, with the highest statistics being used for the lowest-temperature simulations. In these runs of lower temperatures, the annealing events were rarer, so a higher statistics was needed. The number of cases simulated is reflected in the statistical error bars.}

{\red In the current work, the emphasis is the energy release from damage annealing processes,  and the damage production process itself is not our focus. Therefore, we did not include the electronic stopping power \cite{Nor94b} in the simulations. The simulation cell was a cube of silicon atoms with 16 unit cells in each dimension (i.e, a total of $16\times16\times16\times8=32768$ atoms) and periodic boundaries were used in all dimensions.} This cell size was slightly smaller than what {\red had been} normally used in cascade simulations \cite{Nor97f}, due to the necessity to be able to simulate annealing over long time scales {\red at a reasonable computational cost. Although the simulation cell is smaller than the ones used usually in cascade simulations, it is very unlikely for the cascade damage region to touch itself via the periodic boundaries, since in Si the cascade region is broken into many sub-cascade regions when the recoil energy is 5 keV, and the recoil energy can be distributed in the newly formed sub-cascades. To avoid the occurrence of thermally induced defect annealing in cascade simulations, or at least to a great extent alleviate them, we set the ambient temperature of the system to zero by using a Berendsen thermostat on boundary regions to remove heat from the boundaries. The target temperature of this boundary thermostat was 0 K and the temperature damping parameter was $\tau_T = 0.1 ps$.}

In stage (ii), the damage produced in 5 keV cascades was annealed in thermal equilibrium runs over long time scales. {\red Annealing simulations were separated into two big groups. One group utilized PARCAS as the simulation platform, and consequently, this group of annealing simulations only used Berendsen thermostat, since PARCAS only had Berendsen method as the thermostat. All PARCAS annealing simulations in this group lasted for 3 ns. The other group used LAMMPS as the simulation platform. In this group of annealing simulations, some runs were conducted via Berendsen thermostat, while others were carried out using the quantum thermal bath (QTB thermostat). When using LAMMPS, no matter which thermostat we used, QTB or Berendsen, the annealing runs were conducted for 10 ns. The annealing runs were done for temperatures in the range of 0.05-800 K. No matter which simulator we used, PARCAS or LAMMPS, in these annealing runs, the thermostat is always applied to all atoms in the system. }

{\red In the annealing runs, a thermostat was applied to all the atoms. When Berendsen temperature control was used, the temperature damping parameter was $\tau_T = 300 fs$. When using the QTB thermostat, the temperature damping parameter was the same. Besides the thermostat, a Berendsen barostat was also applied to the system for pressure control, regardless of the type of the thermostat. The pressure damping parameter of the barostat was $\tau_P = 3000 fs$. 

The number of annealing runs performed varied, from 7 for the 500 K case to 60 for the 200 K case in the classical simulations, and several hundreds for the quantum thermal bath ones. The highest statistics was used for the lowest temperature cases, since at these temperatures the annealing events were rarer, requiring a higher number of statistics. Another reason for this variation was that the higher temperatures were not of particular interest, and hence a smaller statistics was sufficient for them. The statistics settings in the classical simulations and the quantum thermal bath simulations, namely, the number of cases at each temperature level and each thermostat choice, were reflected in the error bars. The number of recoil cases at each temperature level, interatomic potential, and choice of thermostat, is given in Table \ref{tab:annealing_cases}.}

{\red To enable the analysis in the very next simulation stage on damage annealing events that could occur quite rapidly, the
atom coordinates of all atoms were output every 2 ps throughout the entire simulation. }

\begin{table}[ht]
\centering
\caption{The number of recoil cases used in cascade annealing simulations at different temperatures and thermostats. QTB stands for the quantum thermal bath, and Berendsen stands for the classical Berendsen thermostat. SW stand for Stillinger-Weber interatomic potential, and Tersoff stands for Tersoff III interatomic potential.}
\label{tab:annealing_cases}
\begin{tabular}{|c|c|c|c|}

\hline
Temperature (K) & Thermostat & Number of Cases & Potential \\
\hline
0.05  & QTB        & 800 & SW \\
10    & QTB        & 600 & SW \\
25    & QTB        & 600 & SW \\
50    & QTB        & 400 & SW \\
100   & QTB        & 200 & SW \\
100   & Berendsen  & 200 & SW \\
125   & Berendsen  & 200 & SW \\
150   & QTB        & 200 & SW \\
150   & Berendsen  & 200 & SW \\
200   & QTB        & 200 & SW \\
200   & Berendsen  & 200 & SW \\
300   & QTB        & 200 & SW \\
300   & Berendsen  & 200 & SW \\
300   & Berendsen  & 40 & Tersoff \\
\hline
\end{tabular}
\end{table}

In stage (iii), the atom coordinates output from the annealing
simulations were extracted every 2 ps, each one separately, as input
coordinates for quenching. The quenching simulations were performed
efficiently using Berendsen temperature control \cite{Ber84} towards
0~K while zeroing the velocity of any atom $i$ on which the force
vector $\vec F_i$ was in opposite direction to the velocity
vector $\vec v_i$, i.e. if $\vec F_i \cdot \vec v_i < 0$. {\red This quenching method is called ``quickmin" or ``damped dynamics method" in the manual of LAMMPS \cite{sheppard2008optimization}.}
Using this efficient approach, a quenching time of 1 ps was enough
to get the system temperature down below 0.01 K.
{\red The quenching was done every time when the atom
coordinates were output in the 3 ns annealing runs with Berendsen thermostat, and in the 10 ns annealing runs with QTB thermostats, that is to say, every 2 ps throughout the simulations.
The final potential energy yielded by this energy minimization process was recorded after quenching.}

In stage (iv), the potential energy vs. time in stage (iii) was
analyzed with respect to energy release. {\red After being quenched to 0 K, the energy release events were found to be clearly distinct from random thermal fluctuations in potential energy, that were $< 0.5$ eV. Moreover, when comparing the energy release with the time dependence of the number of Wigner-Seitz defects in the system, one can find out that such energy release could be related to changes in atomic structure.}
To get the total energy release from one {\red annealing} event,
when energy release $> 0.5$ eV occurred on subsequent
2 ps time intervals, the energy release values were summed up as the total release in
the same event. Subsequently, a statistics of the magnitude
of energy release was made. This can
be compared with the experimental observations of energy release, {\red which is the low energy excess signal.}

In stage (v), the annealing time scale was analyzed by fitting a stochastic
time scale law to the data on the time dependence of annealing obtained from
molecular dynamics. {\red To do this fitting, the potential energy curves of all the simulations at each temperature level, from each particular choice of thermostat, Berendsen or QTB, were summed up and averaged, yielding an ``average" potential energy curve at this temperature level. Then, the part of the average curve from 1 ns to the end of the annealing simulations (3 ns for PARCAS simulations and 10 ns for LAMMPS simulations) was selected, and this latter, relatively long-term part, was fit by an exponential function. The data between 0 ns to 1 ns was not used in the time constant fitting, since the most unstable defects are likely to be annealed first, causing an unrealistic amount of energy release.} This allowed us to quantitatively show that annealing
of amorphous pockets can occur {\red on long time scales,} even at extremely low temperatures
relevant for superconducting transition detectors.

\begin{figure}[h]
\begin{center}
\includegraphics[width=0.99\columnwidth]{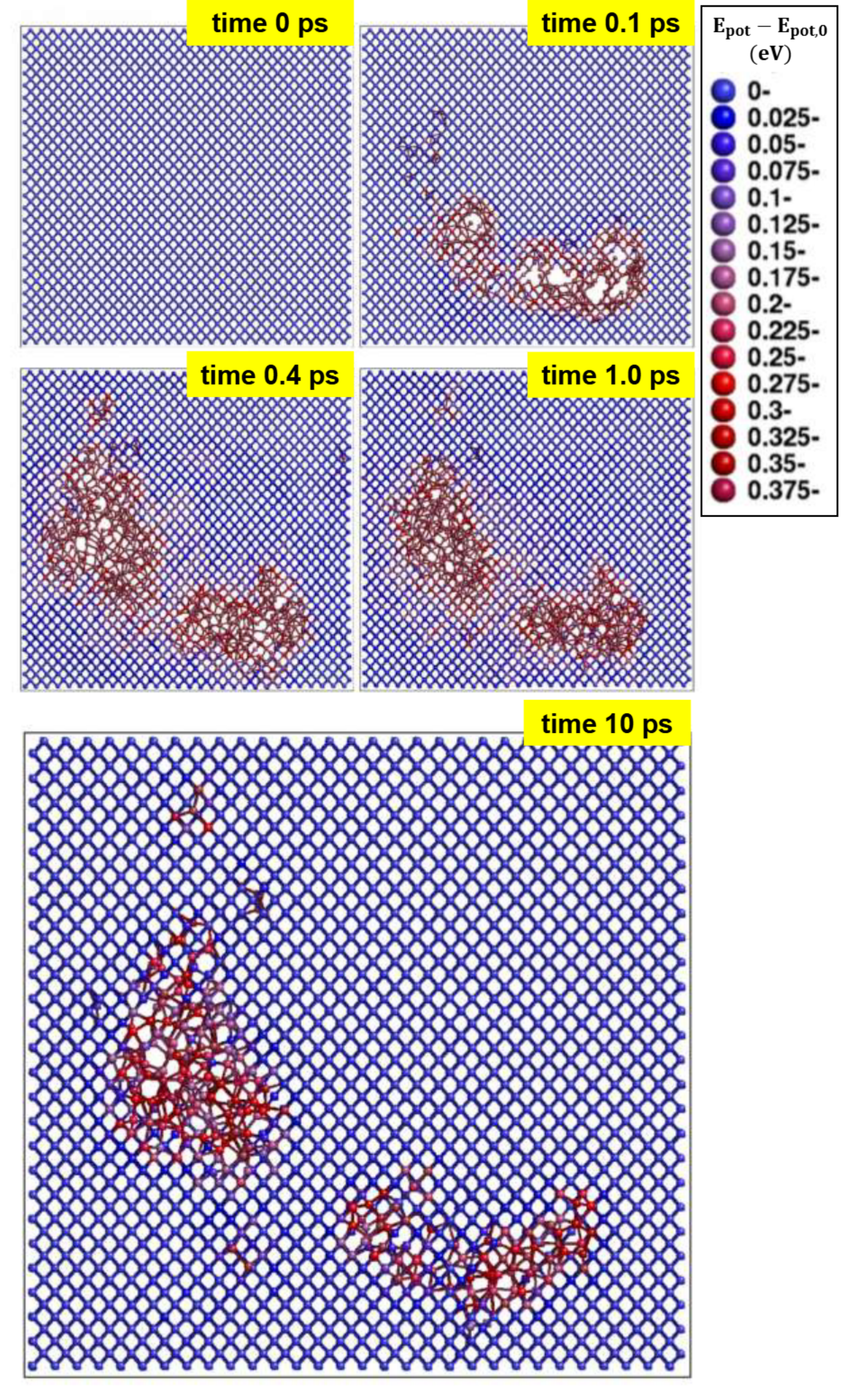}
\end{center}
\caption{\label{fig:cascade}  
Cross-sectional view of the {\red time evolution of damage formation} in a 5 keV cascade in Si, as modeled
with the Stillinger-Weber potential {\red and a classical thermostat. In these simulations, one atom in the cell was given a recoil energy of 5 keV towards the center of the cell, leading to a collision cascade producing damage. For clarity, only atoms in a 2 unit cell thick cross section in the center of the cell are shown. Atoms are of course colliding and damage is produced also in other atom layers than those shown.} The top part shows the {\red time } evolution of the cascade {\red at the specific times 0, 0.1, 0.4 and 1 ps}, and the larger frame at the bottom the final damage state at 10 ps. This cascade was ran at 0 K ambient temperature to ensure there is no thermal annealing. The precise size of the plotted simulation cell region is 81.2 \AA{} in the $x$ (horizontal), 71.3 \AA{} in the $y$ (vertical) and 10.86 \AA{} in the $z$ (out of plane) direction. The colors indicate the potential energy $E_{\rm pot}$ of the atoms relative to the ground state energy $E_{\rm pot,0}$, {\red as indicated in the legend on the top right}.
}
\end{figure}

For defect analysis, the number of defects in the simulation cells was quantified with the
Wigner-Seitz (WS) method \cite{Nor97f}, a geometric method completely independent
of the potential energy analysis. This measure is
used to quantify the number of defects because its definition has no adjustable parameters, and it gives defect numbers roughly consistent with  the simple Kinchin-Pease damage prediction in Si \cite{Nor97f,Nor18}.
However, considering the complex nature of the damage pockets in Figure \ref{fig:cascade}, it is clear that in Si, ``the number of WS defects" does not correspond to any geometrically simple point defect type such as an isolated vacancy or interstitial. Hence we use the term ``the number of WS defect pairs" as the notation of this defect measure. {\red 
 Also, when analyzing the microscopic mechamisms of energy release events, the DXA analysis in Ovito \cite{stukowski2010extracting, stukowski2012automated} was also used to examine the size evolution of the amorphous pockets in silicon, This approach is a quantitative and useful approach for the investigation of the crystallization of amorphous pockets under thermal activation. To use DXA analysis on damaged silicon systems in this project, we utilized the Python package of Ovito, and made programs to calculate the number of atoms in the amorphous phase in the annealing processes. Besides being as handy as WS defect analysis, there is another reason why we used DXA despite the existence of WS defect analysis in our research. In Ovito, DXA analysis can generate a ``defect mesh" to visually mark the boundaries of amorphous pockets in silicon, being very useful when finding out the precise locations of energy release events (in section \ref{sec:location_energy_release}).}

\subsection{Quantum thermal bath simulations}
\label{subsec:quantummd}

To assess how significant quantum mechanical zero point vibrations \cite{Ashcroft-Mermin} were on the annealing time scale and mechanisms, we employed the quantum thermal bath (QTB) approach \cite{Bar11, dammak2009quantum} as implemented in LAMMPS. This QTB thermostat is a Langevin-style thermostat with a specially tuned random force term that agrees with the energy spectrum of quantized vibrations in crystals. This thermostat does not incur very heavy extra computational cost compared with standard molecular dynamics, {\red only roughly doubling the computation time of molecular dynamics simulations}, which is a great advantage compared with other methods, for instance, path integral molecular dynamics \cite{brieuc2016zero}.

When using QTB, the equation of motion is a Langevin style formula, where $f(x)$ is interatomic force, $\gamma$ is the damping constant, 
{\red $t$ is time} and $\Theta$ is the random force term: 
\begin{equation} \label{equation_of_motion}
    m\ddot{x}=f(x)+\sqrt{2m\gamma}\Theta(t)-m\gamma\dot{x}
\end{equation}

The power spectral density of the random force term $\Theta$ is:
\begin{equation} \label{power_spectral_density}
    \Theta(\omega)=\hbar \omega \left[\frac{1}{2}+\frac{1}{\exp \left(\frac{\hbar \omega}{k_B T}\right)-1}\right]
\end{equation}

where $\omega$ is the vibration frequency, $\hbar$ is reduced Planck constant, $k_B$ is Boltzmann constant and T is the temperature of the system. 

In the implementation we use~\cite{Bar11}, the quantum thermal bath has five input parameters, namely, target quantum temperature, temperature damping parameter, a random seed, upper cutoff frequency and the number of frequency bins. To make the results comparable with the results from classical molecular dynamics simulations, we set the temperature damping to be 0.3 ps (300 fs), the same one used in the purely classical simulations. The upper cutoff frequency was set to be three times the Debye frequency of silicon, {\red due to the fact that SW potential overestimated the Debye frequency of silicon \cite{babaei2019machine}.} The number of frequency bins was 100 in all QTB simulations. When using the thermostat, a Berendsen barostat with the settings in section \ref{subsec:classicalmd} was also applied to the system. 

{\red
To check whether these settings enable the compensation of low temperature quantum effect, the heat capacity of pure silicon crystal was calculated by differentiating the mean total energy with respect to temperature (an approach suggested by Dammak et al \cite{dammak2009quantum}), and the results was compared with an experiment \cite{anderson1930heat}. The calculation results from two different thermostats, QTB thermostat and Berendsen thermostat, are compared with each other.}

\begin{figure}[H]
\begin{center} 
\includegraphics[width=1.0\columnwidth]{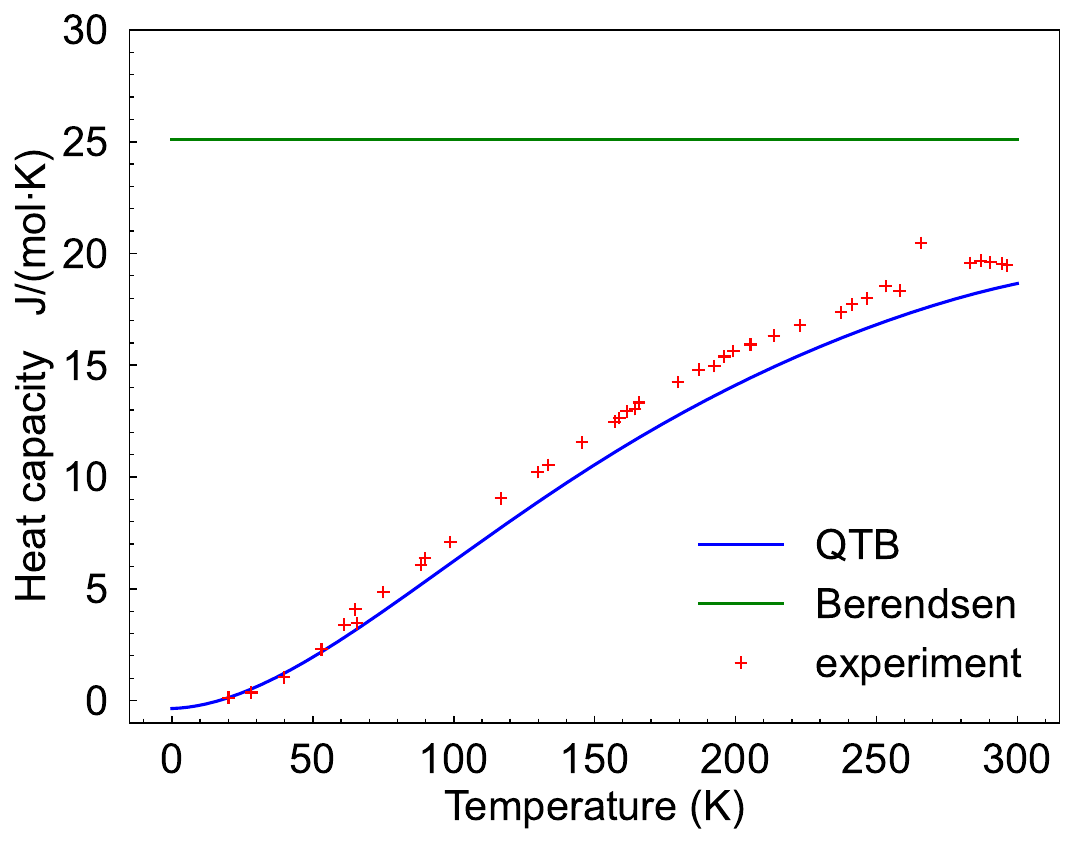} 
\end{center}
\caption{\label{fig:heat_capacity}  
Heat capacity of silicon crystal, {\red calculated with the quantum thermal bath (QTB) and the classical Berendsen thermostat. Since the Berendsen thermostat and molecular dynamics describes only classical atom motion, its heat capacity prediction corresponds to the value given by the equipartition theorem, which does not have a temperature dependence \cite{Mandl}. The results are compared against experimental values (red plus signs). }
}
\end{figure}

{\red The heat capacity curves shown in Figure \ref{fig:heat_capacity} demonstrate that the QTB thermostat can reproduce the temperature dependence of heat capacity at low temperatures. However, with Berendsen thermostat, the calculated heat capacity curve is a completely horizontal line, having a value very close to the heat capacity value given by the Dulong-Petit Law (3R = 24.943 J/(mol$\cdot$K)). This comparison shows that classical thermostats like Berendsen fail in heat capacity calculation, and quantum thermal bath is able to reproduce the temperature dependence of heat capacity, being a quantum effect. 
}

With the help of the quantum thermal bath, the annealing behavior of the system at very low temperatures can be investigated directly in molecular dynamics simulations. The five-stage simulation and analysis procedure in section \ref{subsec:classicalmd} was followed, and quantum thermal bath was used in stage(ii) for the annealing simulations. 

\section{Results and discussions}
\label{sec:results}

\subsection{Stage (i) Damage production in cascade simulations}

An example of a cascade event and the damage produced is shown in a cross-sectional plot in Figure \ref{fig:cascade}. The cross-section shows a very typical damage production event, analyzed in detail in many previous works \cite{Rub95,Nor97f,Nor18}. As noted above, 5 keV is above the sub-cascade threshold breakdown, so the damage is split into separate damage production regions. 1000 different 5 keV cascades were simulated, and the final atom coordinates from these were used as inputs for annealing runs, quenching and energy release analysis.

\subsection{Stage (ii) Annealing simulations}

For the classical molecular dynamics simulations, annealing simulations were performed at several temperatures in the range 20 - 800  K for some of the cells, and at 200 K for all the cells. For the Tersoff potential, annealing was done at 300 K for all the cells. Since the Tersoff potential has a higher melting point than the Stillinger-Weber potential and the experimental one, 300 K corresponds to roughly the same homologous temperature as 200 K for the Stillinger-Weber potential.
For simulations with QTB thermostat, the annealing simulations of 10 ns were conducted at 0.05 K, 10 K, 25 K, 50 K, 100 K, 150 K, 200 K and 300 K. At each temperature, at least 200 annealing simulations with different initial damage were carried out, {\red that is to say, using different final states from different cascade simulations as input structures. Again, the number of annealing cases at each temperature is shown in Table \ref{tab:annealing_cases}.}

\begin{figure}
\begin{center} 
a) \includegraphics[width=1.0\columnwidth]{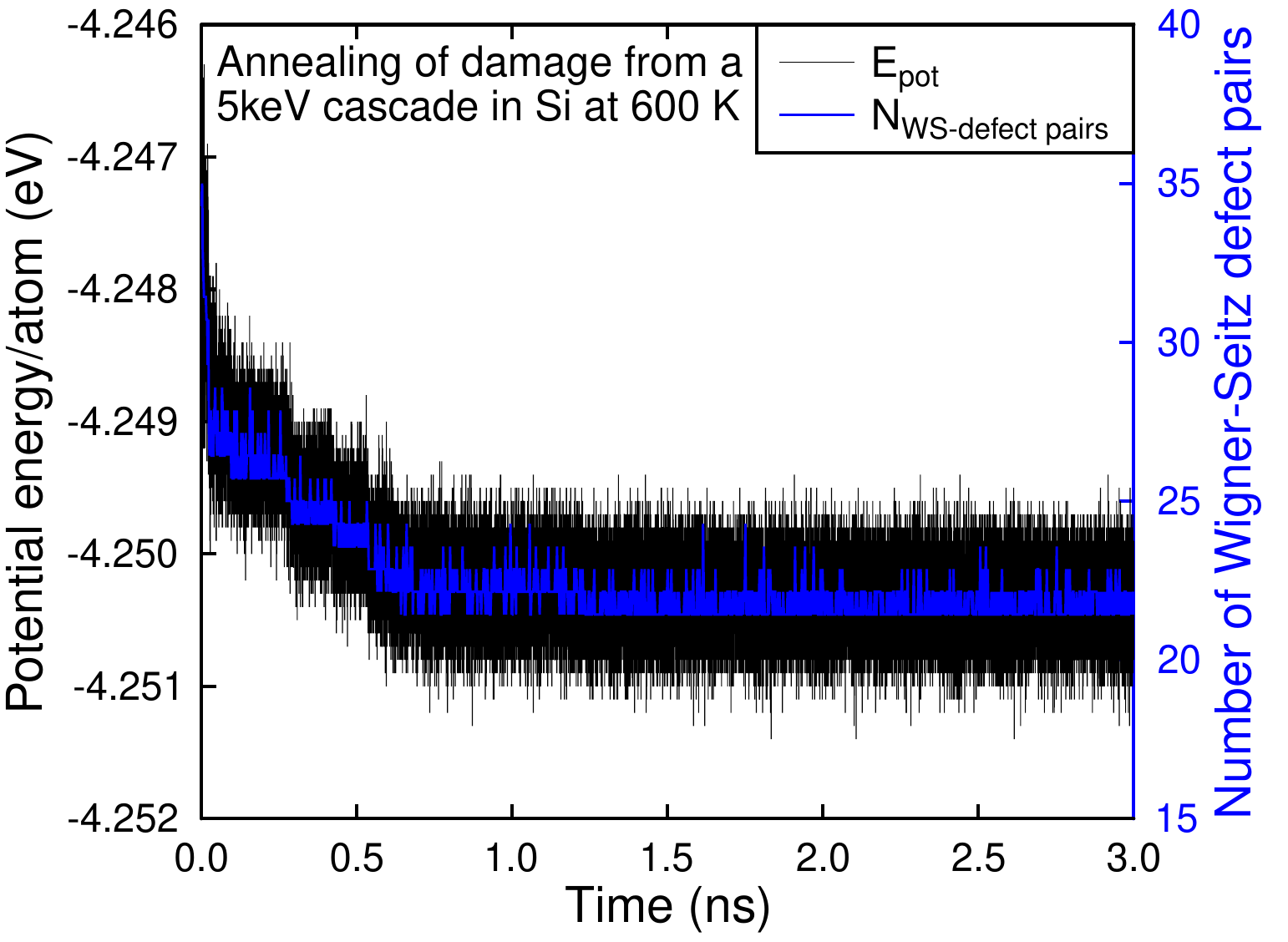} 
b) \includegraphics[width=1.0\columnwidth]{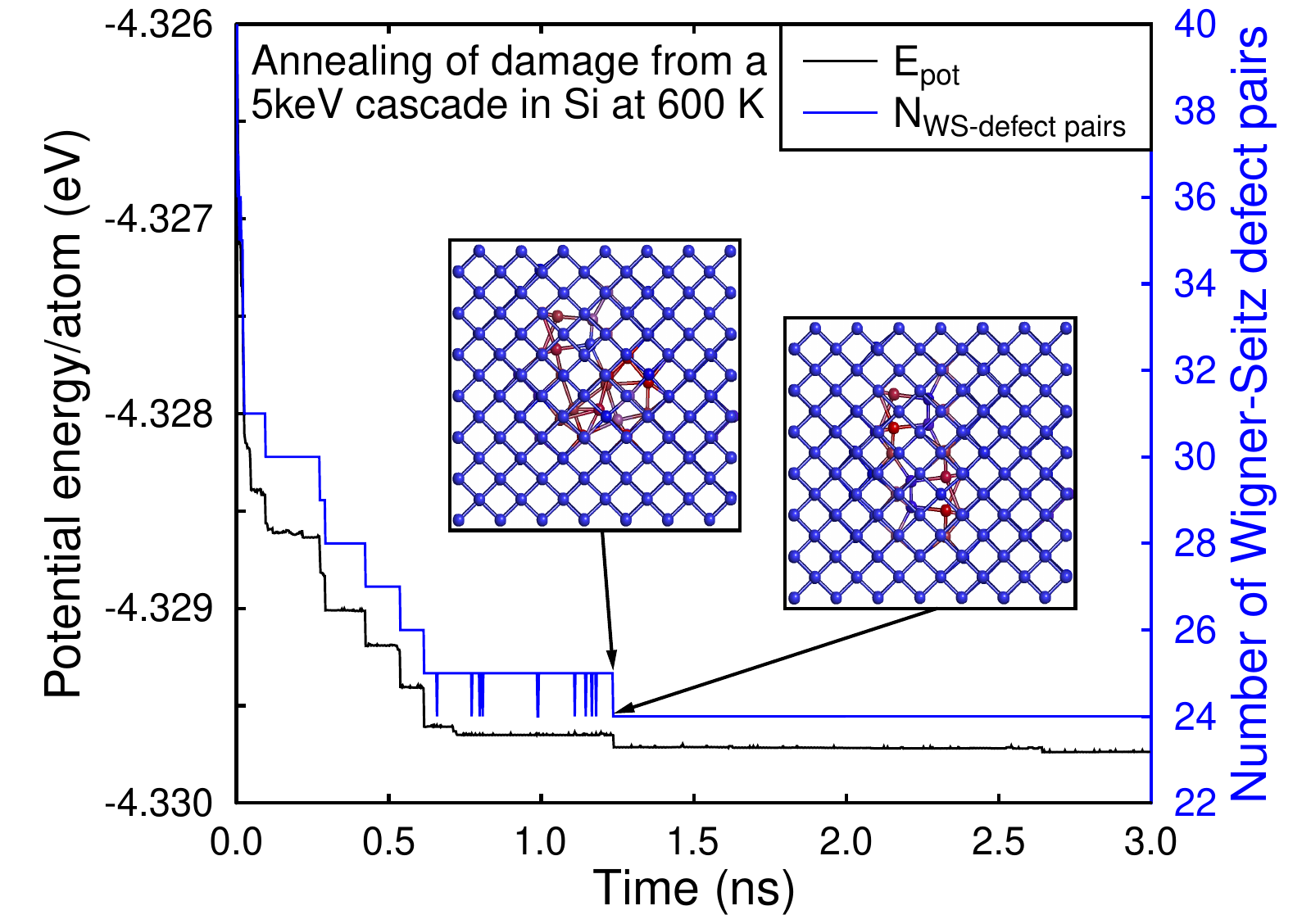} 
c) \includegraphics[width=1.0\columnwidth]{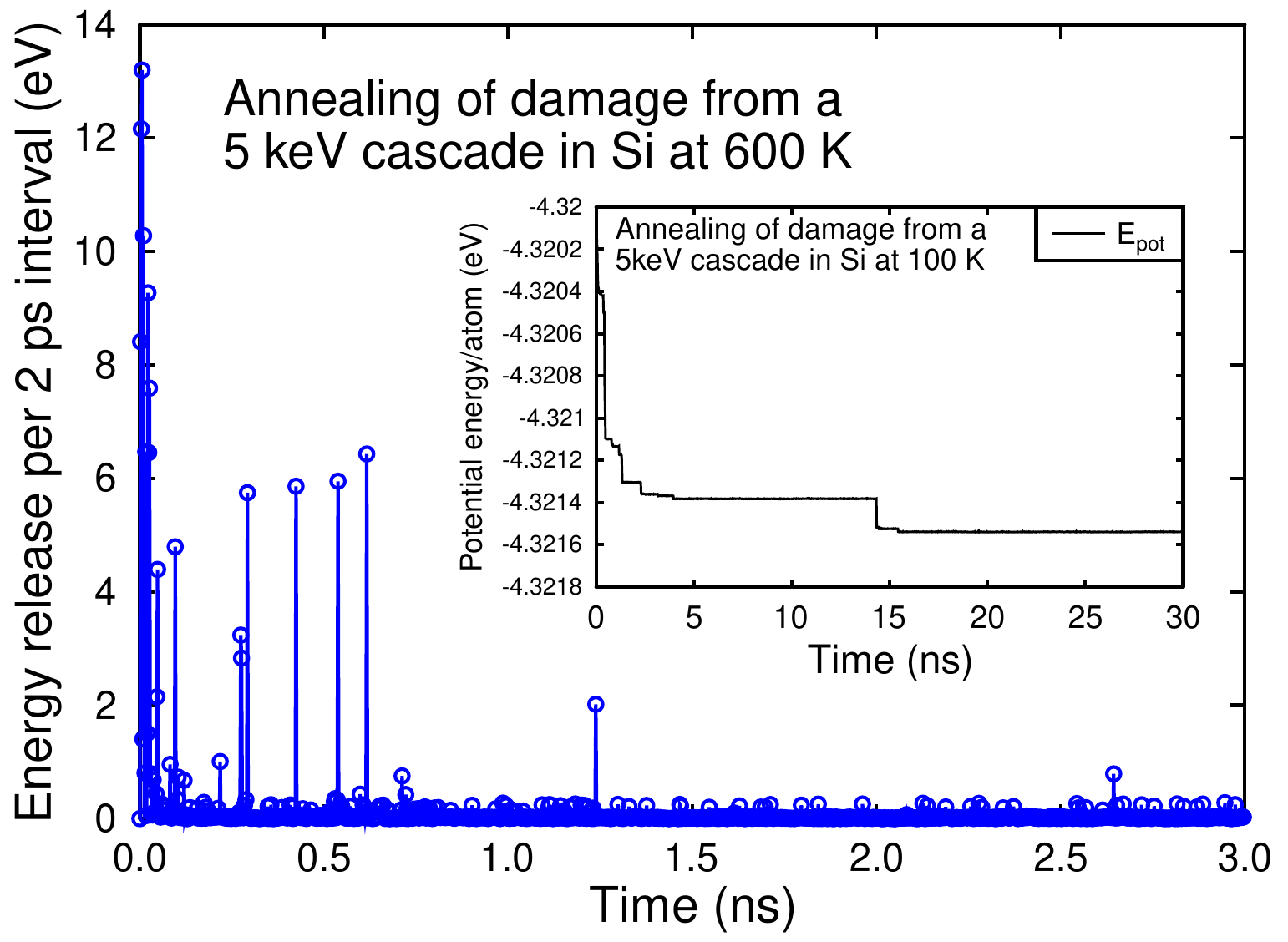} 
\end{center}
\caption{\label{fig:energy_release_time}  
a) Annealing of damage produced in a single 5 keV cascade in Si at 600 K.
The data shows that both the average potential energy and 
the number of Wigner-Seitz defect pairs decreases with time. 
b) Same as subfigure (a) but analyzed after each time step was quenched to 0 K.
The inset shows the atomic positions in one disordered pocket before and after the annealing event occuring at $t=1.236$ ns.
c) Potential energy release per 2 ps intervals from the same cascade.
The inset shows that annealing at well-separated times keep occurring also at longer times and even at the cryogenic temperature of 100 K.
}
\end{figure}

  Results for the number of defects in the cell as a function of time are shown 
in Figure \ref{fig:energy_release_time} (a) for a test 600 K annealing run via classical molecular dynamics. 
The potential
energy has huge fluctuations because a thermodynamic system with a small number of
atoms $N$ has natural fluctuations of the kinetic energy around the average
of the order of $\sqrt{N}/N$ \cite{Mandl}. Since the total energy remains the
same on short time scales, the fluctuations in kinetic energy are also reflected in the
potential energy.
The fluctations make it very difficult to obtain a quantitative analysis
of the potential energy release in a single {\red annealing} event. This was the
reason why we must use the quenching scheme to get rid of thermal
fluctuations.

\subsection{Stage (iii) Quenching runs}
\label{subsec:quenchingresults}

Results of the potential energy and number of WS defect pairs after the quenching
scheme (see section \ref{sec:methods}) are shown in
Figure \ref{fig:energy_release_time} (b), for the exact same annealing run as shown in part \ref{fig:energy_release_time} (a). This graph shows in a much clearer way that after the first about 10 ps, the annealing proceeds stepwise, with clear energy release events occuring every now and then, separated by (on the atomic scale) long times when essentially nothing else than thermal vibrations of atoms around their stable or metastable local equilibrium positions happens in the system. {\red This phenomenon agrees very well with the observations by Caturla et al \cite{caturla1996ion}.}

\begin{figure}
\begin{center} 
\includegraphics[width=1.0\columnwidth]{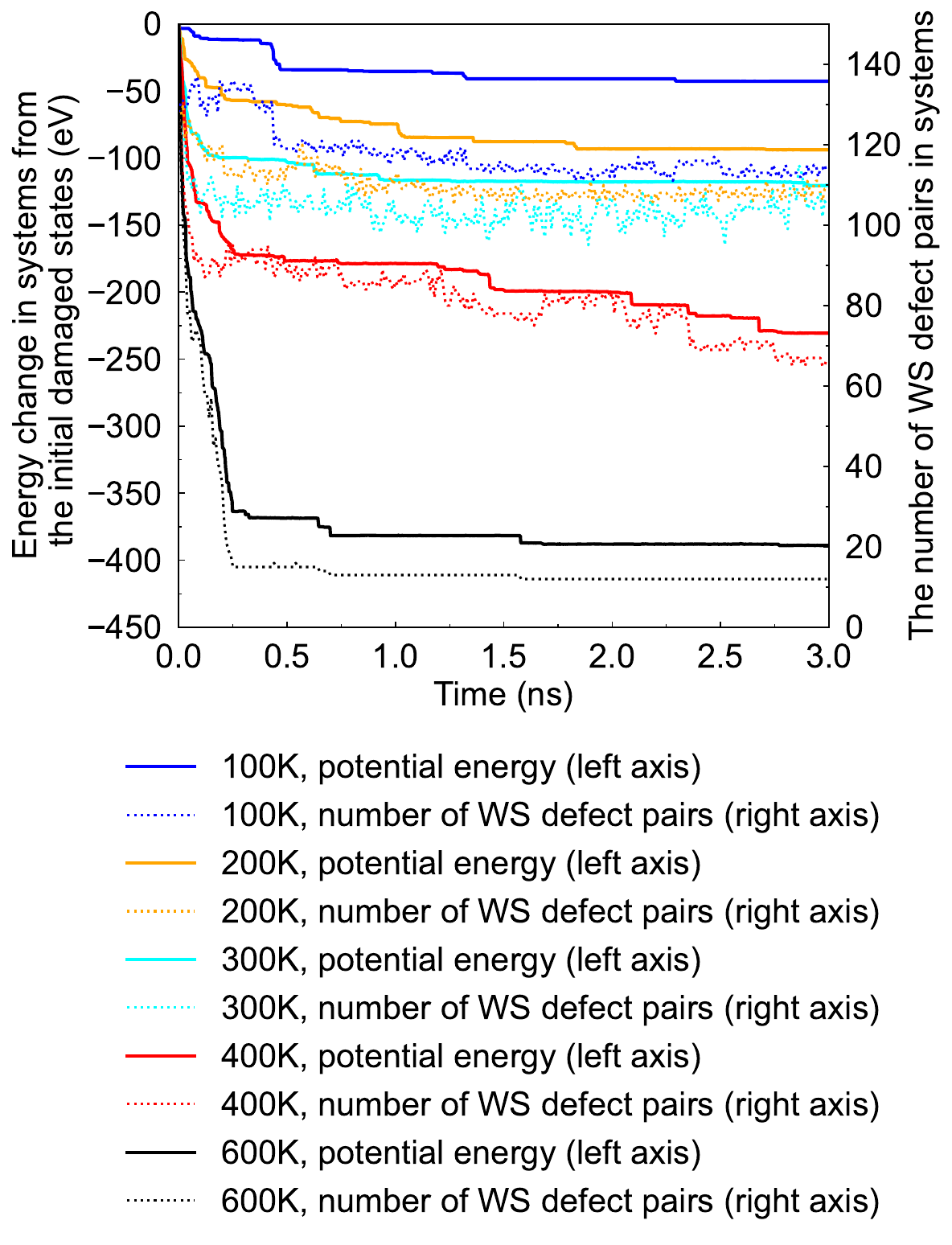} 
\end{center}
\caption{\label{fig:annealing_differentT}  
Annealing of damage produced in a single 5 keV cascade in Si at 
different temperatures.
The data shows that both the average potential energy and 
number of Wigner-Seitz defect pairs decreases with time
at all the considered temperatures. 
}
\end{figure}

{\red Figure \ref{fig:annealing_differentT} demonstrates the potential energy curves and the number of Wigner-Seitz defect pairs together. From this figure, one can easily see that the energy release correlates very well in time with decreases of the number of Wigner-Seitz defects, confirming that the potential energy release is indeed associated with partial annealing of damage in the amorphous pockets, especially the annealing on the amorphous-crystalline interface, where atoms from the amorphous side can rearrange themselves to become crystalline and consequently release energy.}

{\red We note that energy release events may occur either inside the amorphous pockets or at the interface between the crystalline and amorphous phase. There are several reports showing that fully amorphous materials can also undergo relaxation, for example, through the correlated atom motion \cite{Roo92,Oli99,Glo00,Nor04}. Such relaxations would not, however, on average reduce the number of Wigner-Seitz defect pairs, because the atomic structure there is still amorphous after the occurence of energy release events. In our simulations, the energy reduction steps in potential energy correlate almost always in time with a reduction in the number of Wigner-Seitz defect pairs
(Figure \ref{fig:energy_release_time} b and Figure \ref{fig:annealing_differentT}). This indicates that the majority of the large energy release events occur at the interface, which is also observed when the a few selected annealing cases were analyzed in detail (cf. section \ref{sec:location_energy_release}).
}

From the potential energy release, one can also obtain the magnitude of the energy released in each annealing event by comparing the total potential energies of the system before and after each 2 ps annealing interval. This is shown in  Figure \ref{fig:energy_release_time} (c) for the same annealing run as in parts (a) and (b) in Figure \ref{fig:energy_release_time}.
The graph shows that the energy release varies from an annealing event to another.

We ran similar annealing runs for a few cascades in the temperature range 20 -- 800 K. Results for exemplary cases in Figure \ref{fig:annealing_differentT} show that, as expected, at the highest temperatures, very efficient annealing of the damage is observed.
{\red This observation in general agrees with the simulation \cite{Cat96} and the experimental \cite{Lin87,Lin88,Ell88,Hen96} reports showing that it is very difficult to have silicon amorphized by irradiation at temperatures above about 650 K.}

However, we also observe that even at the much lower temperatures, down to 100 K, damage {\red annealing} is observed throughout the 3 ns simulations, {\red even in the cases with the classical Berendsen thermostat.} Also in some test annealing cases for 30 nanoseconds conducted by the Berendsen thermostat, {\red annealing} events kept occurring throughout the entire simulation time: an example of {\red annealing} occurring around 15 ns after the cascade event at 100 K is seen in the inset of Figure \ref{fig:energy_release_time} c. 
Occasional {\red annealing} events are observed in the 50 K and 20 K simulations even in the classical simulations with no quantum thermal vibrations. 
An explanation to this surprising observation is given in section \ref{sec:annmech}. {\red 
These results are well in line with the previous work by B\'eland, Mousseau and co-workers who showed with a hybrid molecular dynamics/barrier search approach (``k-ART") that annealing events can even occur on macroscopic, up to 1 s, time scales \cite{Bel13,Bel15,Jay17}. In the future, we are also going to work on the extension of time scales in our simulation work on defects in dark matter detector materials.
}

\subsection{Stage (iv) Statistics of energy release}
\label{subsec:energy_release_statistics}

To enable a comparison with the experimental observations of the energy
dependence of the low energy excess, we made a statistics of the annealing data similar to that shown in Figure \ref{fig:annealing_differentT}  but summing up all events over at least 50 independent cases (separate cascades as inputs) to obtain a statistically
better picture of the annealing. 

\begin{figure}[h]
\centering
a)
\includegraphics[width=\columnwidth]{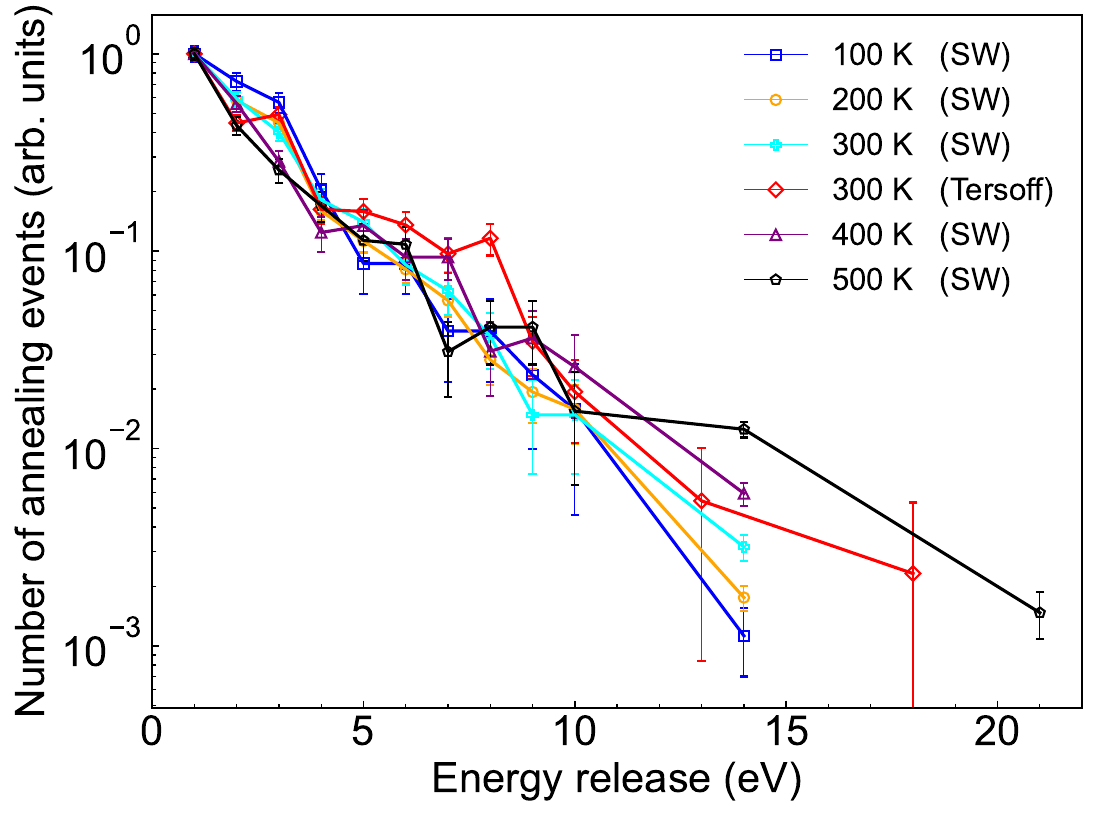}\\
\vspace{1em}
b)
\includegraphics[width=\columnwidth]{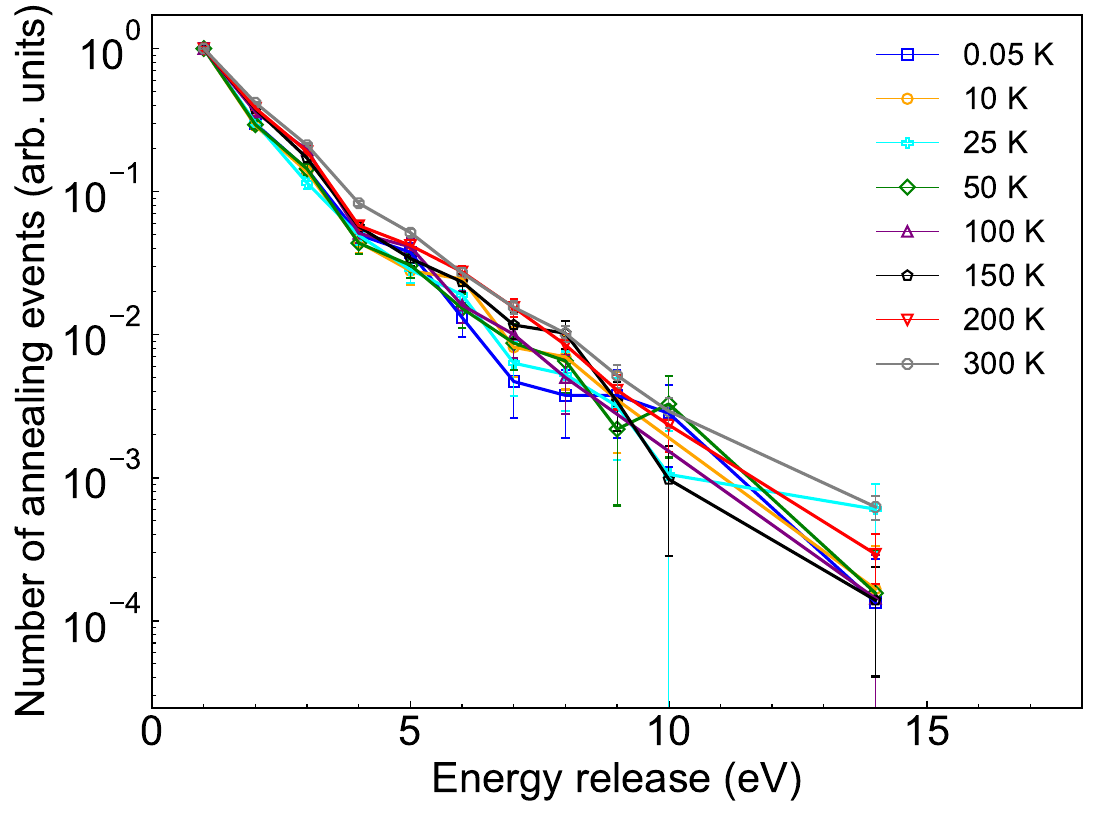}\\
\caption{\label{fig:energy_release_manyT}
Statistics of the magnitude of the energy release events during  annealing runs at different temperatures. To enable easy comparison of the slopes, the curves at different temperatures  have been scaled to having a value of 1 at 1 eV. a) Classical molecular dynamics, b) Molecular dynamics with the quantum thermal bath
}
\end{figure}

For classical molecular dynamics simulations, the results for the temperatures 100 - 500 K are shown in Figure \ref{fig:energy_release_manyT} (a).
Also included in this plot are results obtained with the Tersoff interatomic potential at 300 K.
The results show that, remarkably, the energy release statistics is independent of temperature in the range 100 - 500 K within the statistical uncertainty.   Moreover, the Tersoff interatomic potential gives results practically identical to those of the Stillinger-Weber potential. Since Tersoff and Stillinger-Weber potential are developed completely independent of each other, having different functional forms, and being fitted to different properties \cite{Ter88c,Sti85}, this gives high confidence that the obtained results are not an interatomic potential artifact. 

For simulations with QTB themostat, the results at various temperatures are shown in Figure \ref{fig:energy_release_manyT} (b). {\red It is very clear that the energy release statistics are independent of temperature, since the magnitudes of the slopes are hardly related to the changes in temperature.} 

In both Figure \ref{fig:energy_release_manyT} (a) and Figure \ref{fig:energy_release_manyT} (b), the data are clearly close to linear on the log-lin plot, showing that the energy release statistics follows an exponential dependence with energy release magnitudes, in agreement with the experimental observations \cite{CRESST:2019axx,CRESST:2019jnq,DAMIC:2020cut,EDELWEISS:2019vjv,EDELWEISS:2020fxc,CRESST:2017ues,NUCLEUS:2019kxv,SENSEI:2020dpa,SuperCDMS:2018mne,SuperCDMS:2020ymb}. As expected from the linear dependence of the plots, the data can be fit very well with the functional form:

\begin{equation} \label{exponential_fitting}
    f(E) = A\exp (-\alpha E)
\end{equation}

where $E$ is the energy release and $A$ and $\alpha$ are fitting constants. The fits were done with the Levenberg-Marquardt approach for least square fitting of arbitrary functional forms \cite{Bevington,NumericalRecipes}.
The same functional form has been used earlier to fit the low-energy tail
of the experimentally observed energy releases \cite{Abbamonte:2022rfh,Hei22}.
Results for the fits of the slope coefficient $\alpha$ are given in Table \ref{tab:expfits}.
Considering the statistical uncertainties, one can conclude that the slope of the energy release depends either very weakly or not at all on temperature.

\begin{figure}
\begin{center} 
\includegraphics[width=1.0\columnwidth]{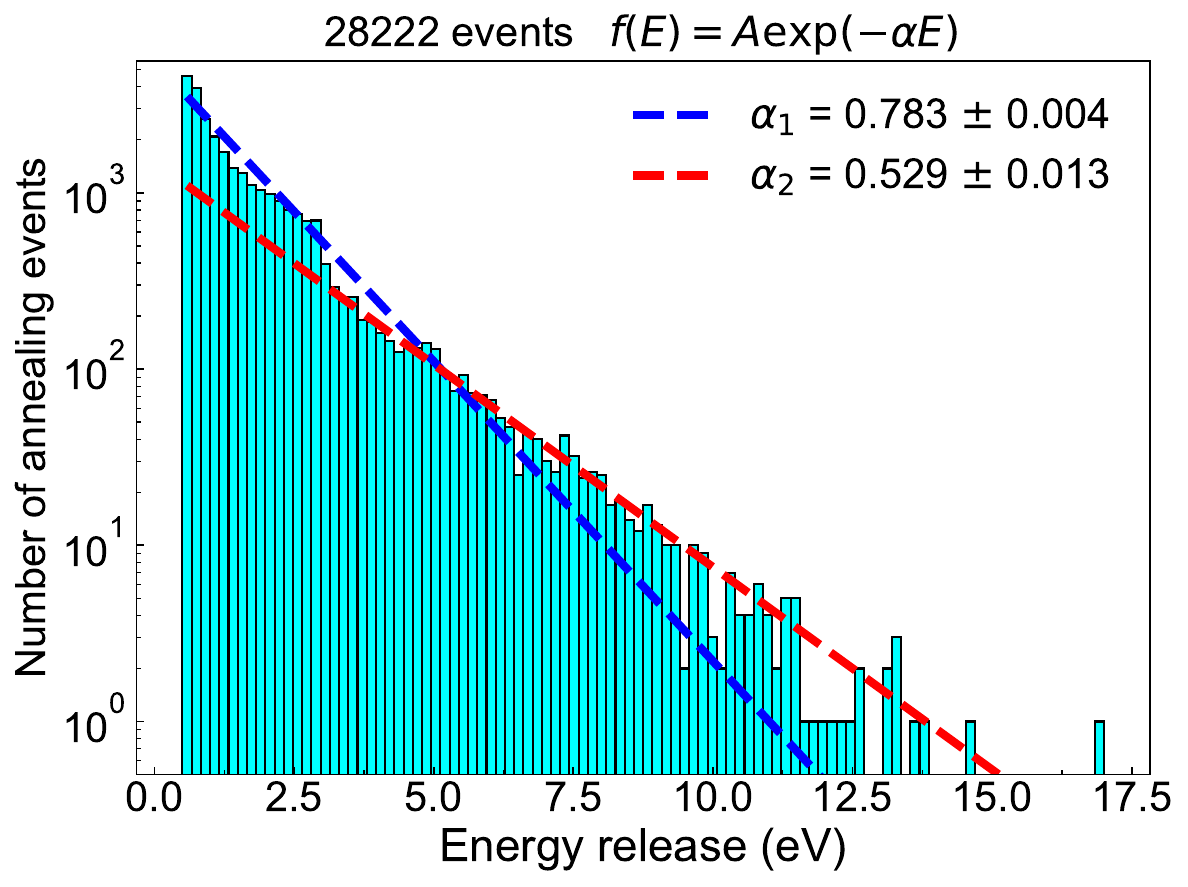}  
\end{center}
\caption{\label{fig:energy_release_all_temperature}  
{\red The histogram of all the energy release events from all temperature levels. The energy release events from annealing simulations at 0.05 K, 10 K, 25 K, 50 K, 100 K, 150 K, 200 K, 300 K are gathered together to obtain an improved energy release statistics. In total, 28222 events are collected.}
}
\end{figure}

{\red
Since the slope coefficient $\alpha$ is barely related to temperature, we can make an attempt to gather together all the energy release events at all the temperature levels (0.05 K, 10 K, 25 K, 50 K, 100 K, 150 K, 200 K, 300 K) in simulations with QTB thermostat, and plot a combined histogram of the energy release statistics of all the annealing events. This combined histogram is Figure \ref{fig:energy_release_all_temperature}. In this figure, 100 histogram bins are used, and the energy release spectrum can be fitted by exponential functions. However, it is worth noticing that we can obtain two slope coefficients by choosing the energy release range for fitting. The first slope $\alpha_1$ is calculated by fitting the entire histogram, while the second slope $\alpha_2$ is obtained by only fitting the energy range larger than 4.0 eV. We can clearly see that $\alpha_2$ is significantly smaller than $\alpha_1$. Compared with $\alpha_1$, we attach much more importance to $\alpha_2$, which in terms of energy is closer to experimental energy release spectra and can consequently be compared against experimental results. 
}

\begin{figure}
\begin{center} 
\includegraphics[width=1.0\columnwidth]{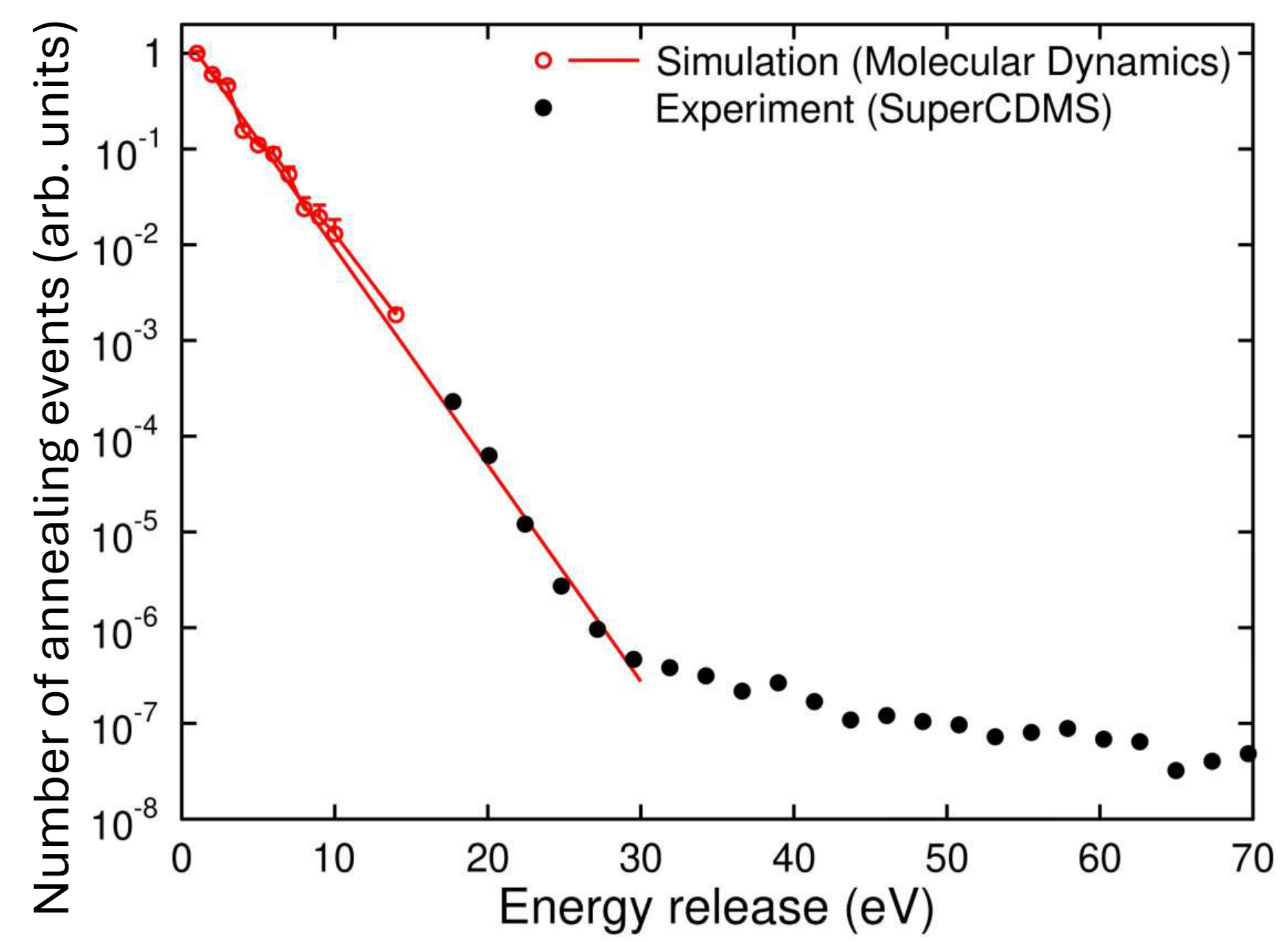}  
\end{center}
\caption{\label{fig:energy_release_expt}  
Comparison of the energy dependence of energy release events with experiments.
Note that the red line is a fit of an exponential form to the simulation data
only, showing that the slope of the low-energy signal from simulations
agrees excellently with experiments. The experimental data is from
Ref. \cite{SuperCDMS:2020aus}.
}
\end{figure}

The simulated and experimental data from a silicon detector are compared to each other
in Figure \ref{fig:energy_release_expt}. The $y$ axis scale is 
in arbitrary units, since in experiments the absolute values of the signal depends on factors such as detector size, detection efficiency, concentration of radioactive impurities creating the initial damage, and damage recoil spectrum, all of which are not fully well known. However, the functional dependence of the signal on energy can be used for a direct comparison between simulation and experiments.

The Figure \ref{fig:energy_release_expt} shows a very good agreement: both the simulation and low-energy tail of the experimental data clearly follow an exponential trend with a very similar slope. 
The slope obtained from a fit to the simulation data at 0.05 K (QTB thermostat) was $\alpha_{\rm sim} = 0.58 \pm 0.05$ 1/eV, which is in perfect agreement with a fit of the same exponential law to the low-energy tail of the experimental data \cite{Hei22}
$\alpha_{\rm exp} = 0.57 \pm 0.01$ 1/eV (to enable direct comparison, this value was obtained by fitting the experimental data with the same software as used to fit the simulated ones). We emphasize that no fitting of the simulations to the experiments has been used for obtaining this outstanding agreement. 
{\red 
Also, the slope coefficient in our combined histogram (Figure \ref{fig:energy_release_all_temperature}) is around 0.53 in the high energy regime, being very close to the experimental value. Based on very similar slope values, Figure \ref{fig:energy_release_manyT}, Figure \ref{fig:energy_release_all_temperature} and Figure \ref{fig:energy_release_expt} together can indicate that the energy release events brought about by the defect evolution in detector materials can possibly be the source of the low energy excess signals in dark matter detectors. }

\begin{table}
\caption{\label{tab:expfits}
Slope coefficients $\alpha$ of the exponential dependence $\exp(-\alpha E)$ of the energy
release $E$ from simulations and experiments on Si. The simulation results obtained with the classical Berendsen thermostat are marked ``Berendsen". and the ones with the quantum thermal bath ``QTB".
}
\begin{tabular}{cccc}
   & T (K) & Type &  $\alpha$ (1/eV) 	\\
\hline
Simulation          & 500  & Berendsen   & 0.49 $\pm$ 0.03 \\
(current work)      & 400  & Berendsen  & 0.49 $\pm$ 0.03  \\
                   & 300   & Berendsen & 0.53 $\pm$ 0.03  \\
                    & 300 & QTB & 0.52 $\pm$ 
                    0.01 \\
                   & 200   & Berendsen & 0.52 $\pm$ 0.02  \\
                   & 200   & QTB & 0.51 $\pm$ 
                   0.02  \\
                   & 100   & Berendsen & 0.60 $\pm$ 0.05  \\
                   & 100   & QTB & 0.58 $\pm$
                   0.07  \\
                   & 50   & QTB & 0.52 $\pm$
                   0.06  \\
                   & 25   & QTB & 0.49 $\pm$
                   0.04  \\
                   & 10   & QTB & 0.50 $\pm$
                   0.06  \\
                   & 0.05  & QTB & 0.58 $\pm$ 
                   0.05   \\
Experiment              & 0.04  & --    & 0.57 $\pm$ 0.01 \\
\hline
\end{tabular}
\end{table}

\subsection{Stage (v) Analysis of annealing time scales}

We finally analyze the time dependence of the annealing events
 such as those shown in Figure \ref{fig:energy_release_time} and \ref{fig:annealing_differentT}.
To be able to roughly estimate the annealing time scales as a function
of temperature and extrapolate to even lower temperatures, we analyzed
the time dependence of the annealing curves. 

\begin{figure}
\begin{center} 
\includegraphics[width=1.0\columnwidth]{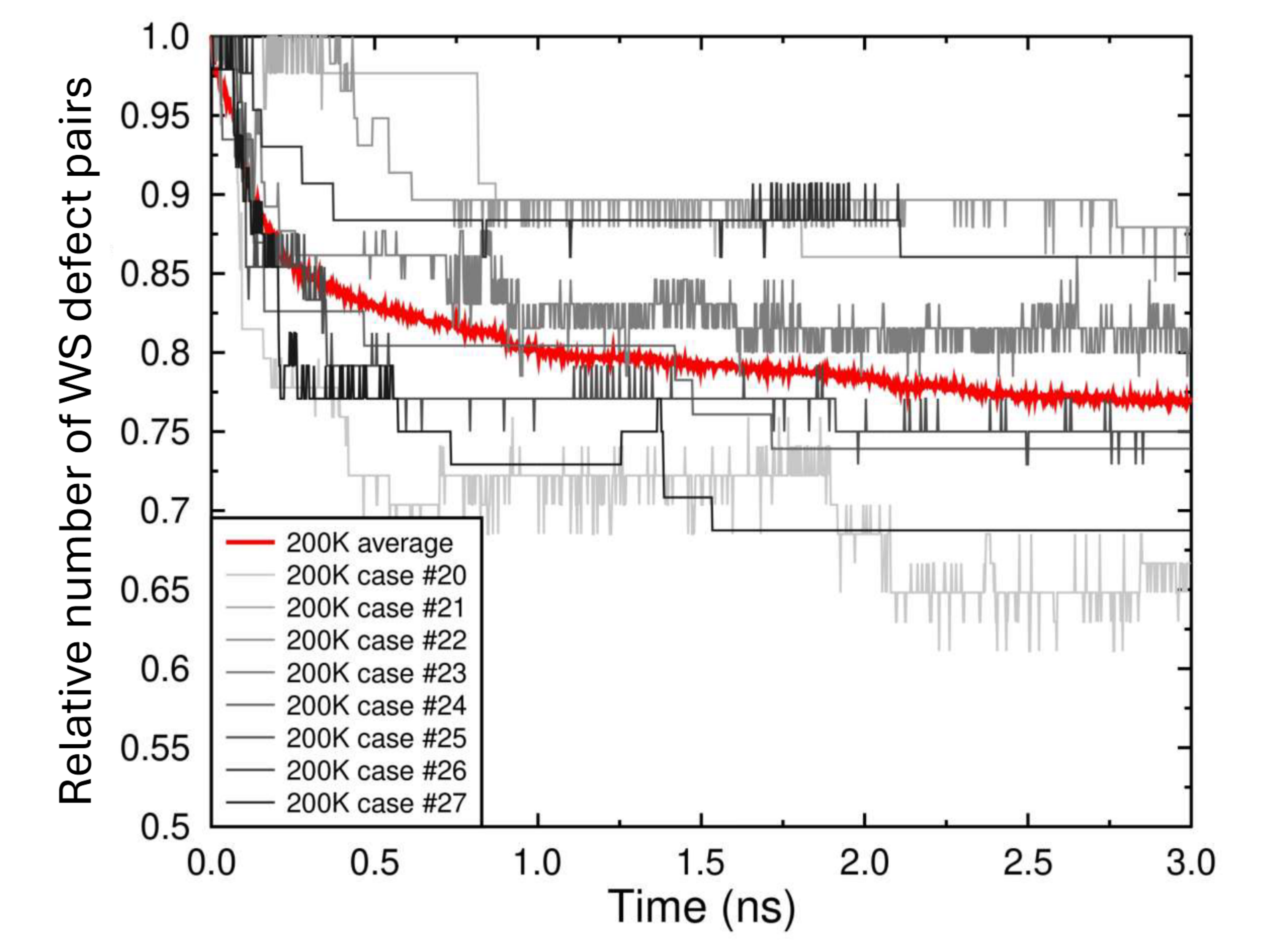} 
\end{center}
\caption{\label{fig:defectfraction_variations}  
Annealing of damage produced on average in several 5 keV cascades in Si at 
200 K plotted in terms of the fraction of Wigner-Seitz defects
compared to the initial state in 7 individual cases (shades of gray) and the average over 50 cases (red). Not All individual cases are shown since it is difficult
to read fifty shades of gray.
}
\end{figure}

\begin{figure}
\begin{center} 
a) 
\includegraphics[width=1.0\columnwidth]{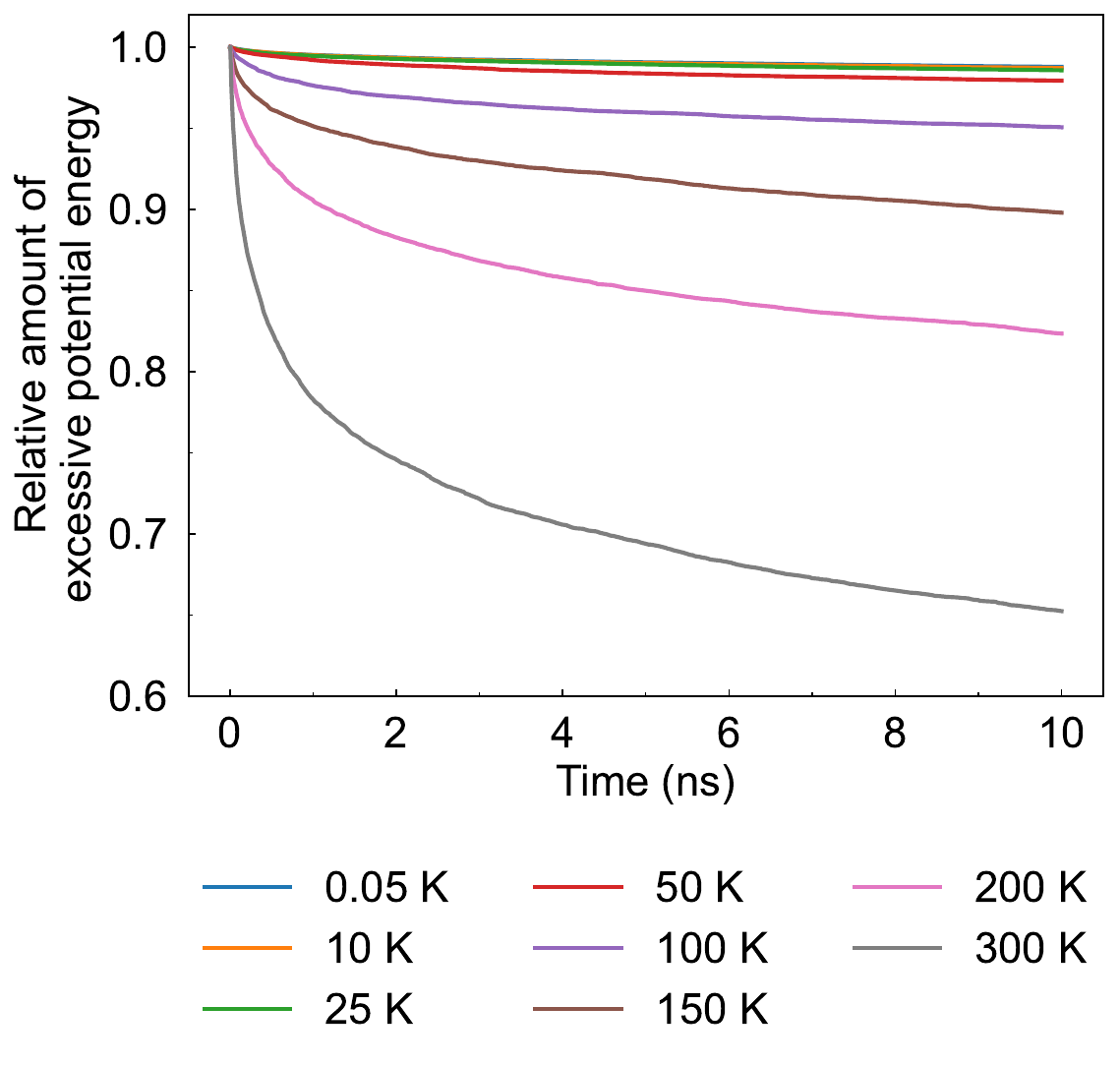} \\
\vspace{1em}
b) 
\includegraphics[width=1.0\columnwidth]{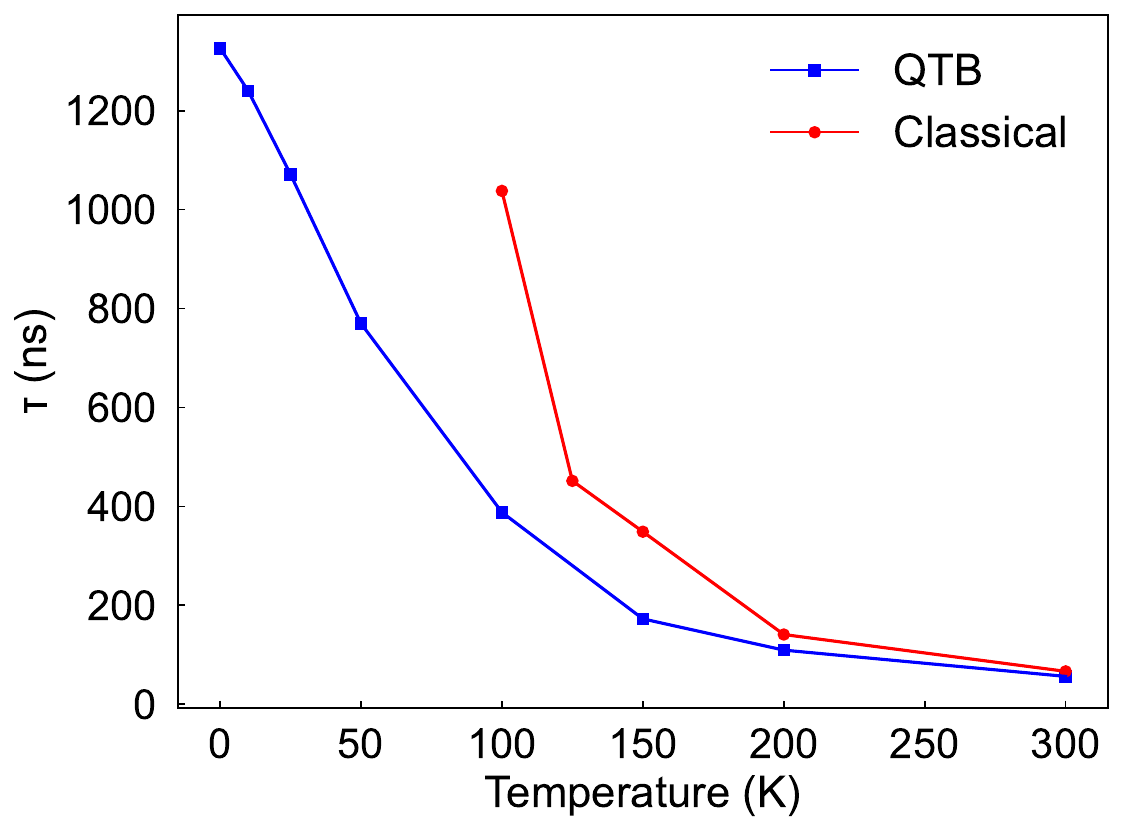} 
\end{center}
\caption{\label{fig:annealing_curves}  
a) Average potential energy curves with the quantum thermal bath after quenching. The values are rescaled so that the initial value is one. In this way, the curves show the relative amount of excessive potential energy. 
b) {\red Results of fits of the annealing time parameter $\tau$ to data on the time dependence of potential energy in annealing, being a function of the annealing temperature. Since in the classical simulations practically no annealing processes are observed below 100 K, it was not possible to provide an annealing time constant for the lowest temperatures in classical cases.}
}
\end{figure}

Since the number of disordered pockets in a single event is small, the annealing curves of single events have huge statistical
fluctuations. This is illustrated in Figure \ref{fig:defectfraction_variations}. 
To obtain a statistically reliable analysis, we carried out an analysis of at least 200 different cascade events {\red at each temperature level} in the temperature range 0.05 -- 300 K. To be able to compare cases with different numbers of initial defects and stored energy in them, the analysis was done for the average
relative fraction of the excessive potential energy remaining at time $t$ compared to the initial excessive potential energy at $t=0$. The fraction $f$ was calculated for each cascade event separately before averaging. The results 
of the averaged fraction analysis are shown in Figure \ref{fig:annealing_curves} (a). As expected, time evolution of the fraction of remaining Wigner-Seitz defects showed a similar behaviour.

The averaged annealing curves show that the annealing behaviour can be divided in two stages: a rapid annealing occurring roughly during the first 0.5 ns after the damage event, and a subsequent much slower annealing after this.
The first stage clearly is due to {\red annealing} of the most metastable defects, whereas the later {\red annealing} is an average over occasional well-separated 
events, corresponding to the late energy release 
events shown in Figure \ref{fig:energy_release_time}.

By comparing the individual cases shown in Figure \ref{fig:defectfraction_variations}, an additional important point is illustrated: the damage in some events, especially those with large damage pockets, can anneal much slower than the average energy release. On long time scales, annealing from such cases would dominate the residual recovery signal.

\begin{table}
\caption{\label{tab:fitting_time_parameter}
Time parameter $\tau$ in the fitting $A\exp(-t/\tau)$ of the potential energy curves. The error bar is the uncertainty estimate of the parameter obtained from the least squares fitting procedure. The results obtained by classical molecular dynamics with Berendsen thermostat are marked ``Berendsen" and the ones with the quantum thermal bath ``QTB".
}

\begin{tabular}{cccc}
   & T (K) & Type &  $\tau$ (ns) 	\\
\hline
& 300   & Berendsen & 66.38 $\pm$ 0.29 \\
& 300 & QTB & 56.19 $\pm$ 0.20 \\
& 200   & Berendsen & 140.86 $\pm$ 0.47  \\
& 200   & Qtb. & 109.33 $\pm$ 0.41  \\
& 150   & Berendsen & 348.8 $\pm$ 1.4  \\
& 150   & QTB & 172.53 $\pm$ 0.46  \\
& 125   & Berendsen & 451.5 $\pm$ 1.8  \\
& 100   & Berendsen & 1037.8 $\pm$ 3.1  \\
& 100   & QTB & 388.0 $\pm$ 1.3  \\
& 50   & QTB & 769.8 $\pm$ 2.5  \\
& 25   & QTB & 1070.9 $\pm$ 2.4  \\
& 10   & QTB & 1234.0 $\pm$ 3.3  \\
& 0.05  & QTB & 1326.1 $\pm$ 3.7   \\
\hline
\end{tabular}
\end{table}

The curves in Figure \ref{fig:annealing_curves} were fitted against the exponential function $A\exp(-t/\tau)$. In the fitting practice, the initial 1 ns in the potential energy curves was truncated, and the rest of the curves were fitted. The fitting results and the comparison with the results from purely classical annealing are shown in Figure \ref{fig:annealing_curves} (b) and Table \ref{tab:fitting_time_parameter}.

{\red We note that in previous works \cite{Bel15}, a roughly logarithmic time dependence, namely, a relaxation being linear with log(time) in annealing, was reported for damage in Fe. In the current case, we found that at room temperature, the time dependence was indeed logarithmic, but at the lower temperatures it was not, at least up to the time scale of 10 ns studied here. We also note that due to limited numbers of annealing events at the lower temperatures, the statistics is limited and we cannot be certain that the time dependence on even longer times can be described with the function $\exp(-t/\tau)$ with a single time constant. However, the current fits to a slowly decaying time dependence are sufficient to show that the randomly timed annealing events do occur up to at least millisecond time scales after the initial picosecond time scale radiation event. Hence the annealing energy release events may appear uncorrelated to the initial radiation production event.
}

From these results, it is very clear that the low temperature quantum effect can facilitate annealing significantly. 
{\red This is because the quantum zero point energy brought by the  quantum bath 
may trigger the atomic rearrangements as they only need to overcome barriers around $\lesssim 0.1$ eV (cf. Figs. \ref{fig:annealing_event1} (b) and \ref{fig:annealing_event2} (b)), which are comparable to the zero point energy $E_{\rm zp} \sim 0.06$ eV, using the Debye model estimate $E_{\rm zp} = 9/8 k_B \Theta_D$ and Debye temperature of $\Theta_D \approx 630$ K for silicon \cite{Ashcroft-Mermin,silicon_debye_temperature}. 
At the lowest temperatures, these energy barriers are not accessible for the thermal vibrations in simulations with classical thermostats, since the classical thermal energy is $k_B T \ll 0.1$ eV when $T \lesssim 100$ K.
}

The simulation at 0.05 K with QTB thermostat shows that the annealing time parameter $\tau$ has a magnitude of microseconds at the operation temperature of the detectors. The fact that at 10 K and 0.05 K time constants are almost the same indicates that at temperatures $\lesssim 10$ K, the triggering of annealing effects is dominated by the quantum effect. 

The exponential decay of annealing effects could be detectable if the time resolution of the experiments is shorter than the time constants $\tau$. In contrast, if time resolution is much longer than this, the energy release events would seem to appear at random times with average constant rate, which would be determined by the concentration of radioactive impurities that produce the initial damage.

\section{Avalanche mechanism of low-temperature annealing}
\label{sec:annmech}

The observation described in subsection \ref{subsec:quenchingresults} and subsection \ref{subsec:energy_release_statistics} of annealing occuring at cryogenic temperatures
may seem surprising, because the temperature of 100 K corresponds to only roughly 0.01 eV/atom, much lower than the chemical bond energy of roughly 2 eV/bond. It is also not consistent with the traditional picture of radiation damage annealing, which is usually described to be caused by point defect migration with well-defined migration barriers. 

However, the result can qualitatively be understood as follows. The complex disordered damage pockets (cf. Figure \ref{fig:cascade}) can inadvertently have metastable bonds with a much lower energy barrier to {\red annealing}. Furthermore, once a metastable 
bond recombines towards the ground state energy, it can induce
a localized {\red annealing} ``chain reaction'': one bond starts annealing, then the released energy heats up the sample
locally which can induce additional bond {\red rearrangements}. To illustrate this argument with a rough calculation, even a low energy release  of 2 eV of one recombining bond corresponds to
a transient heating of the nearest ten atoms by about 1000 K. To quantify this deduction, we now analyze in atomic-level detail what kind of mechanisms can lead to the energy releases of several eV's even when the temperature is so low that thermal activation cannot possibly lead to crossing an energy barrier in the eV range. 

To do the analysis, some of the the 2 ps intervals where energy release events took place were picked out. Annealing simulations were restarted at the beginning of these intervals, and atomic information was recorded every 10 fs. The atomic configuration extracted every 10 fs became the input for quenching. Quenching was performed in the same way as before (Section \ref{sec:methods}) to find out energy minima along the reaction path. Finally, transition state search by climbing image nudged elastic band (CI-NEB) method was carried out between neighboring energy minima to find out the energy barriers.

\begin{figure*}
\newpage
\begin{center} 
\begin{tabular}{ll}
a) 
\includegraphics[width=0.95\columnwidth]{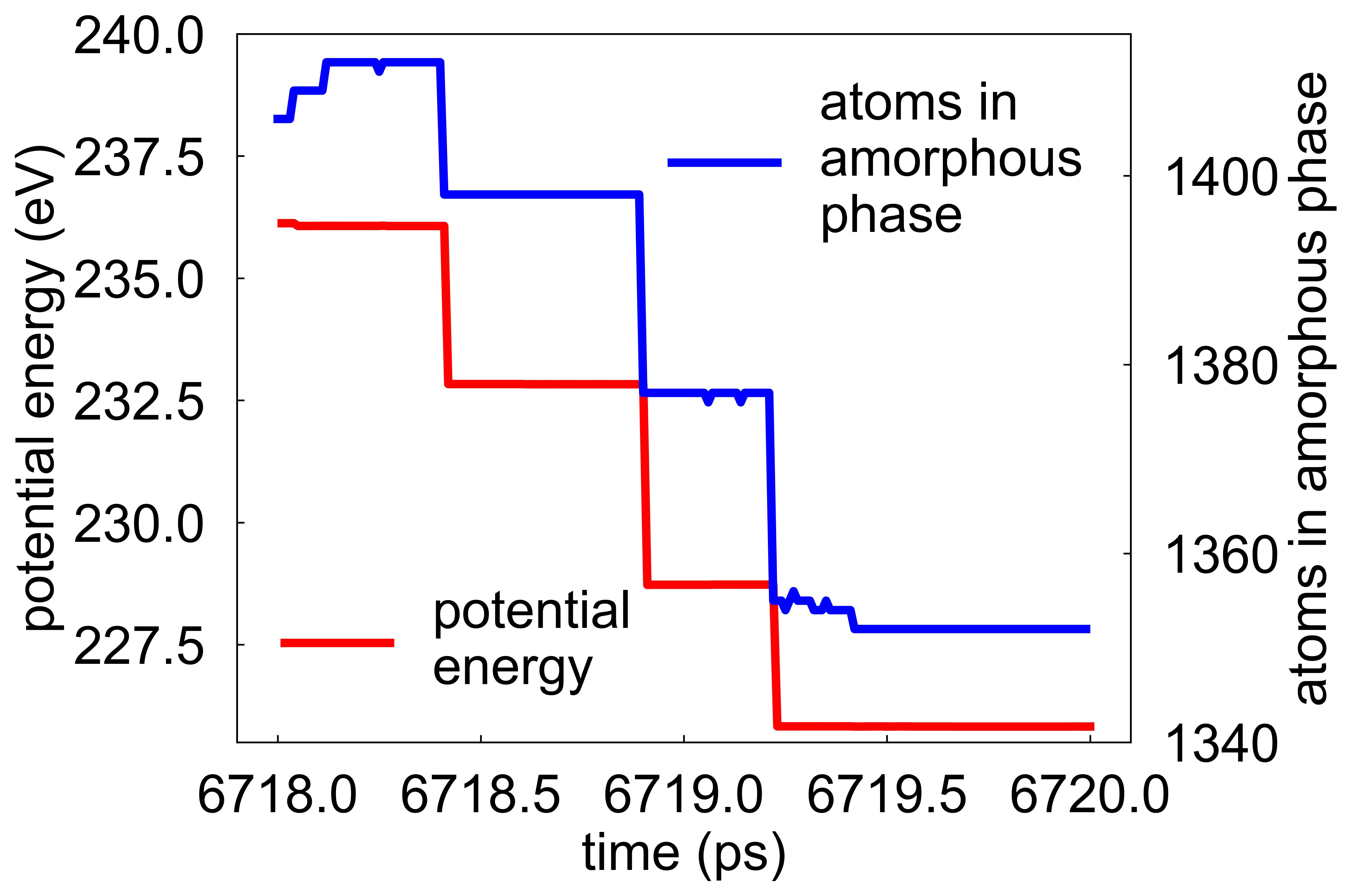} 
&
b) 
\includegraphics[width=0.95\columnwidth]{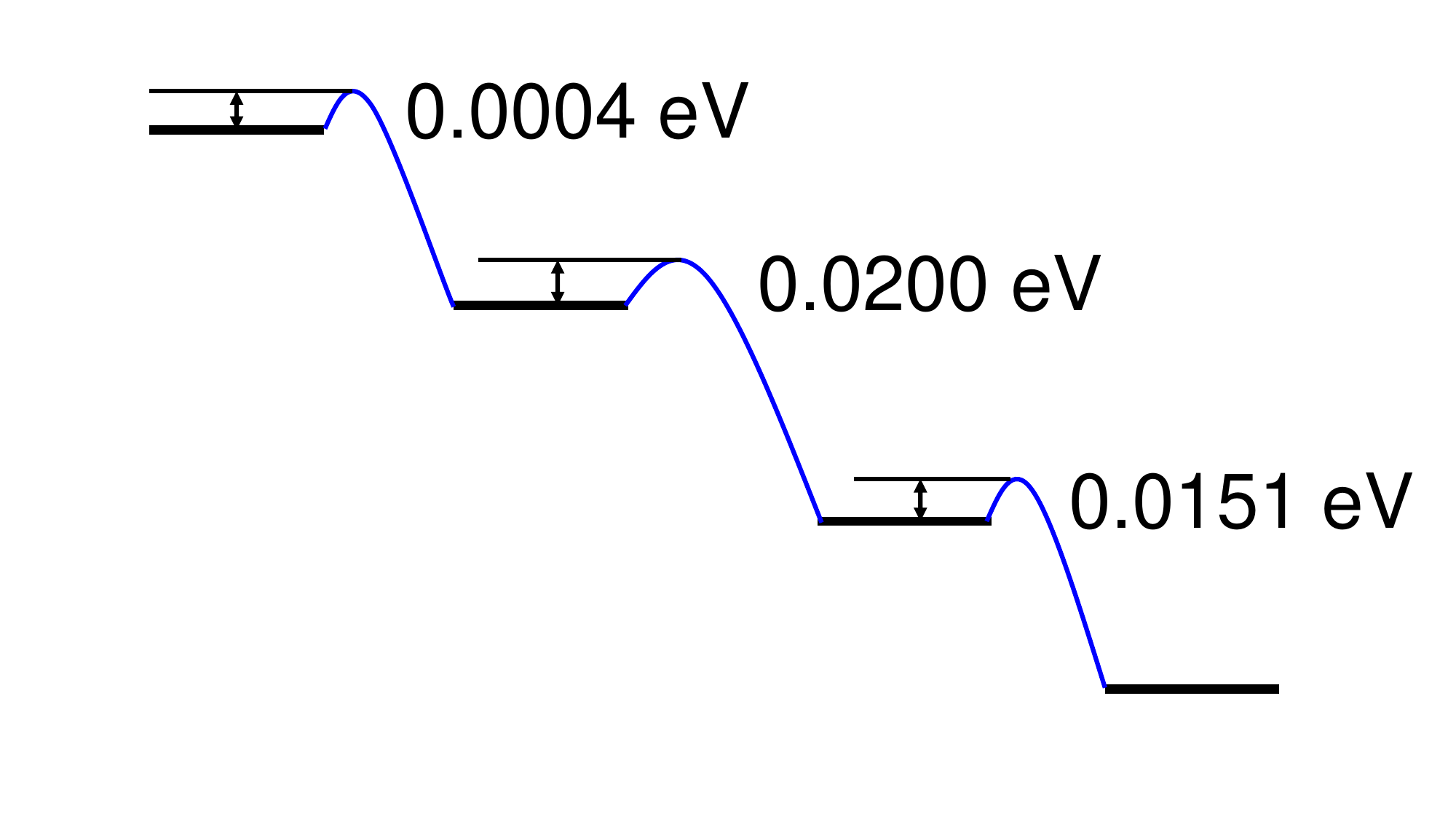} 
\\
c) 
\includegraphics[width=0.95\columnwidth]{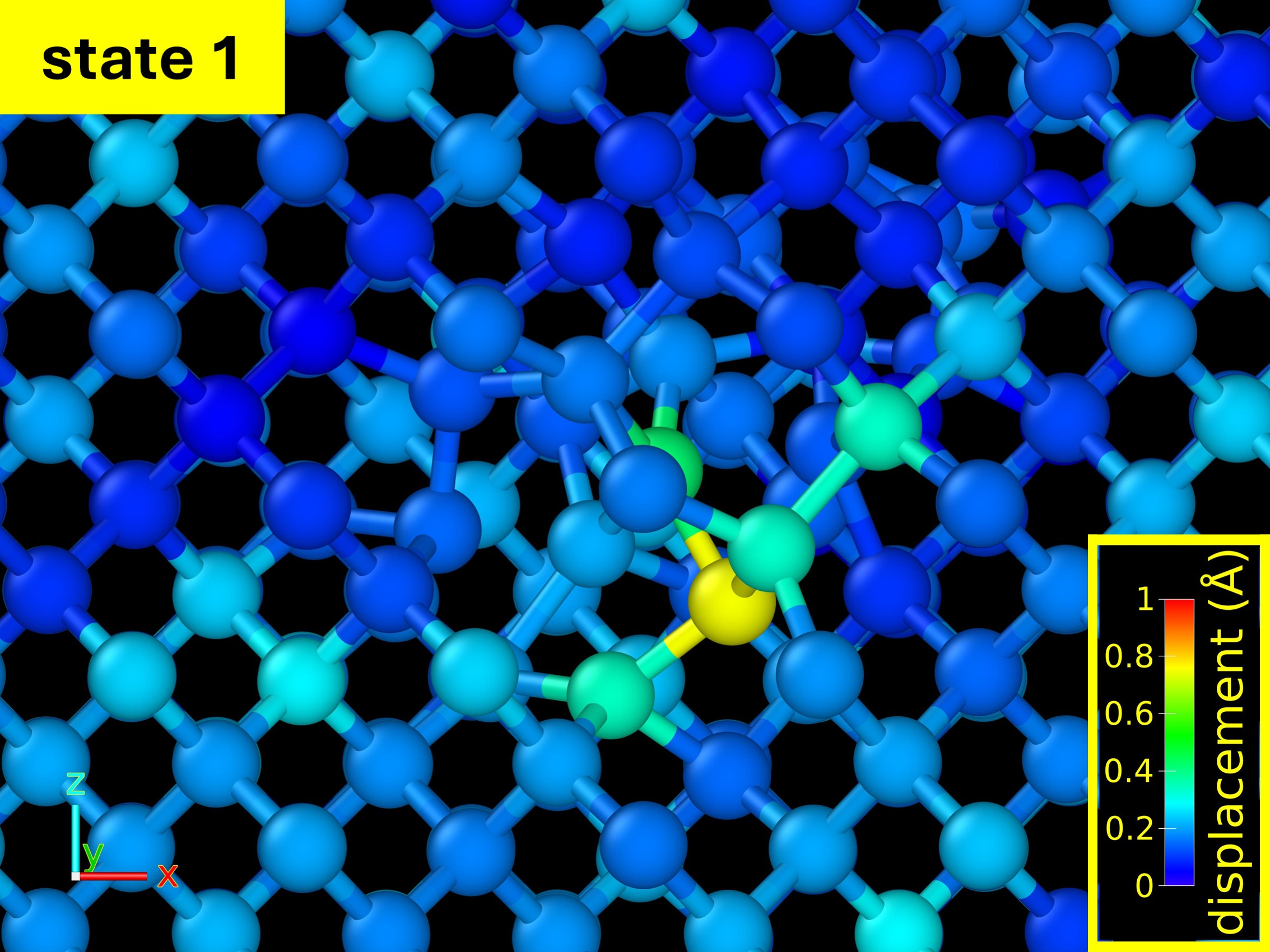} 
&
d)
\includegraphics[width=0.95\columnwidth]{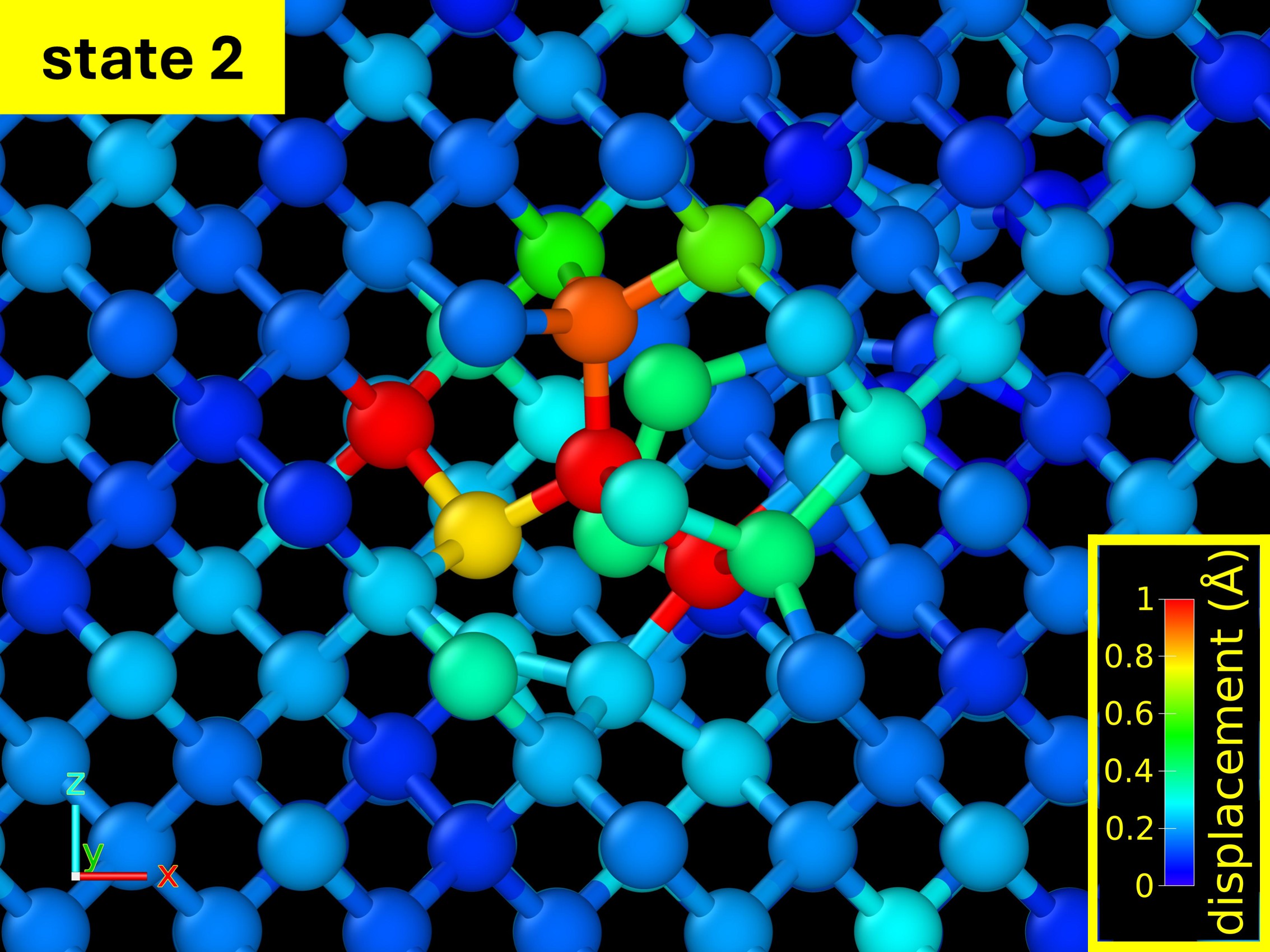} 
\\
e)
\includegraphics[width=0.95\columnwidth]{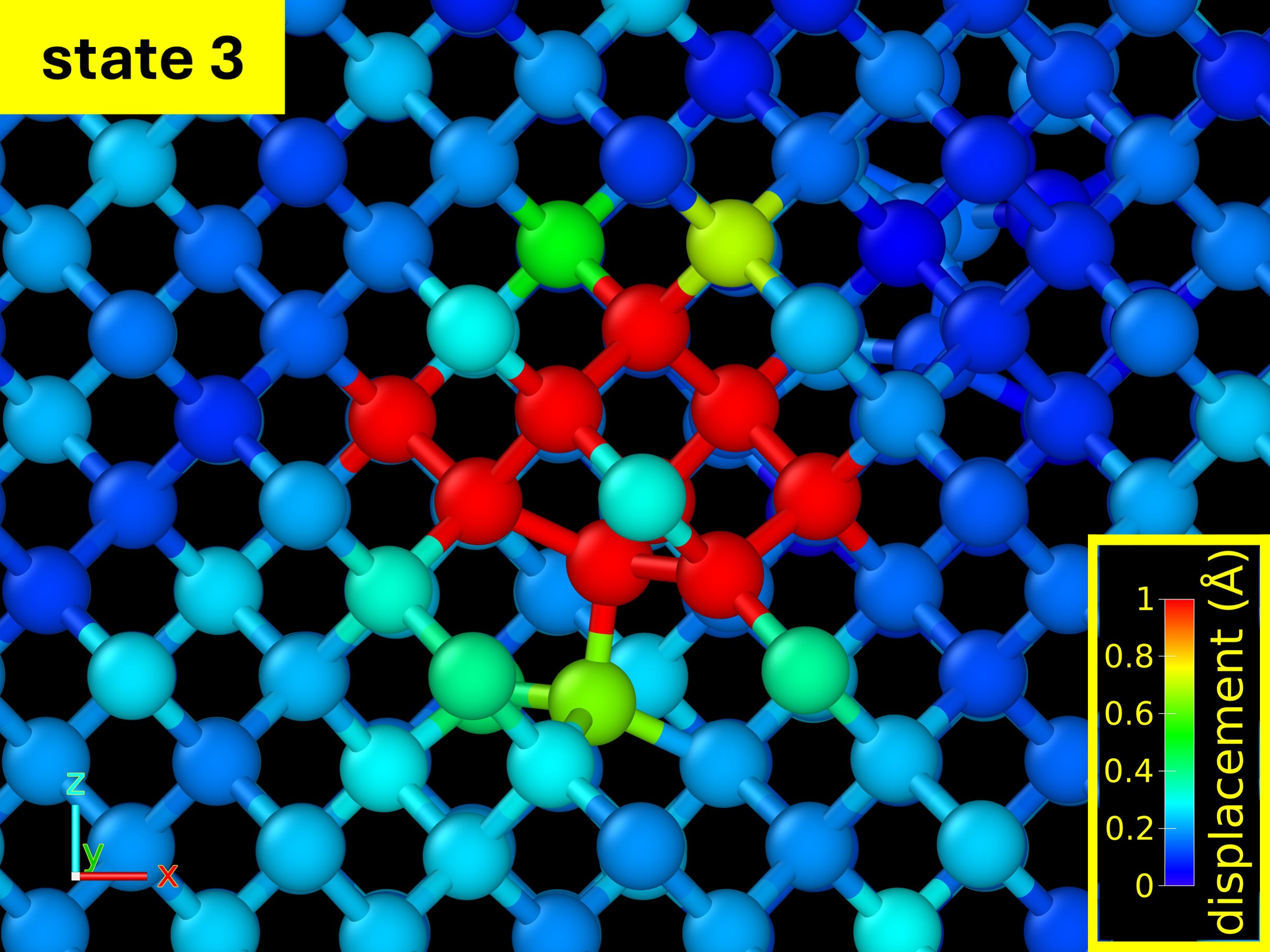} 
&
f)
\includegraphics[width=0.95\columnwidth]{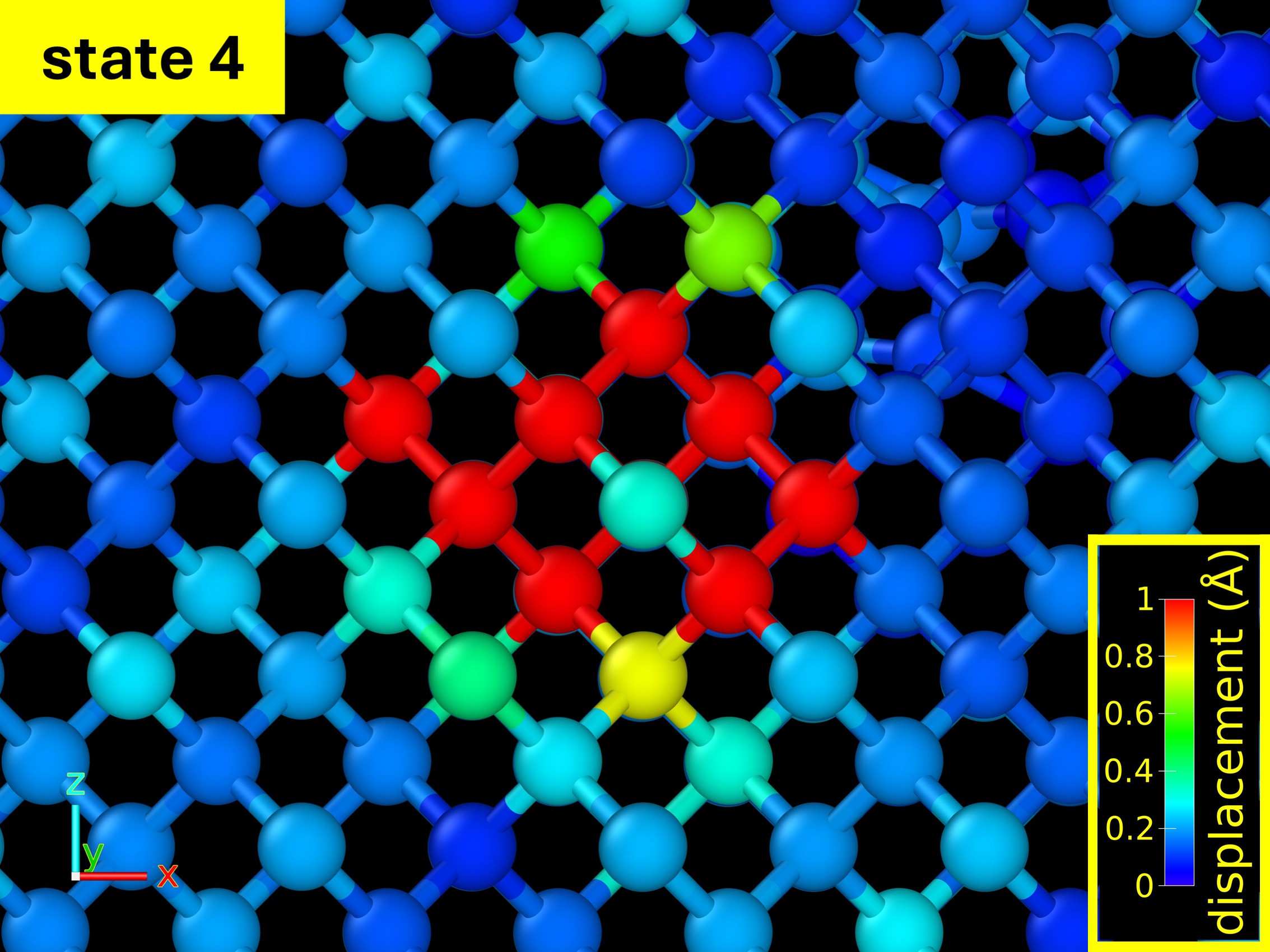} 
\\
\end{tabular}
\end{center}
\caption{\label{fig:annealing_event1}  
An example of the energy barriers in an individual annealing event. In this case the total energy release was 10.24 eV. {\red In (c)(d)(e)(f), the color coding is the atomic displacement. }
}
\end{figure*}

\begin{figure*}
\newpage
\begin{center} 
\begin{tabular}{ll}
a) 
\includegraphics[width=0.95\columnwidth]{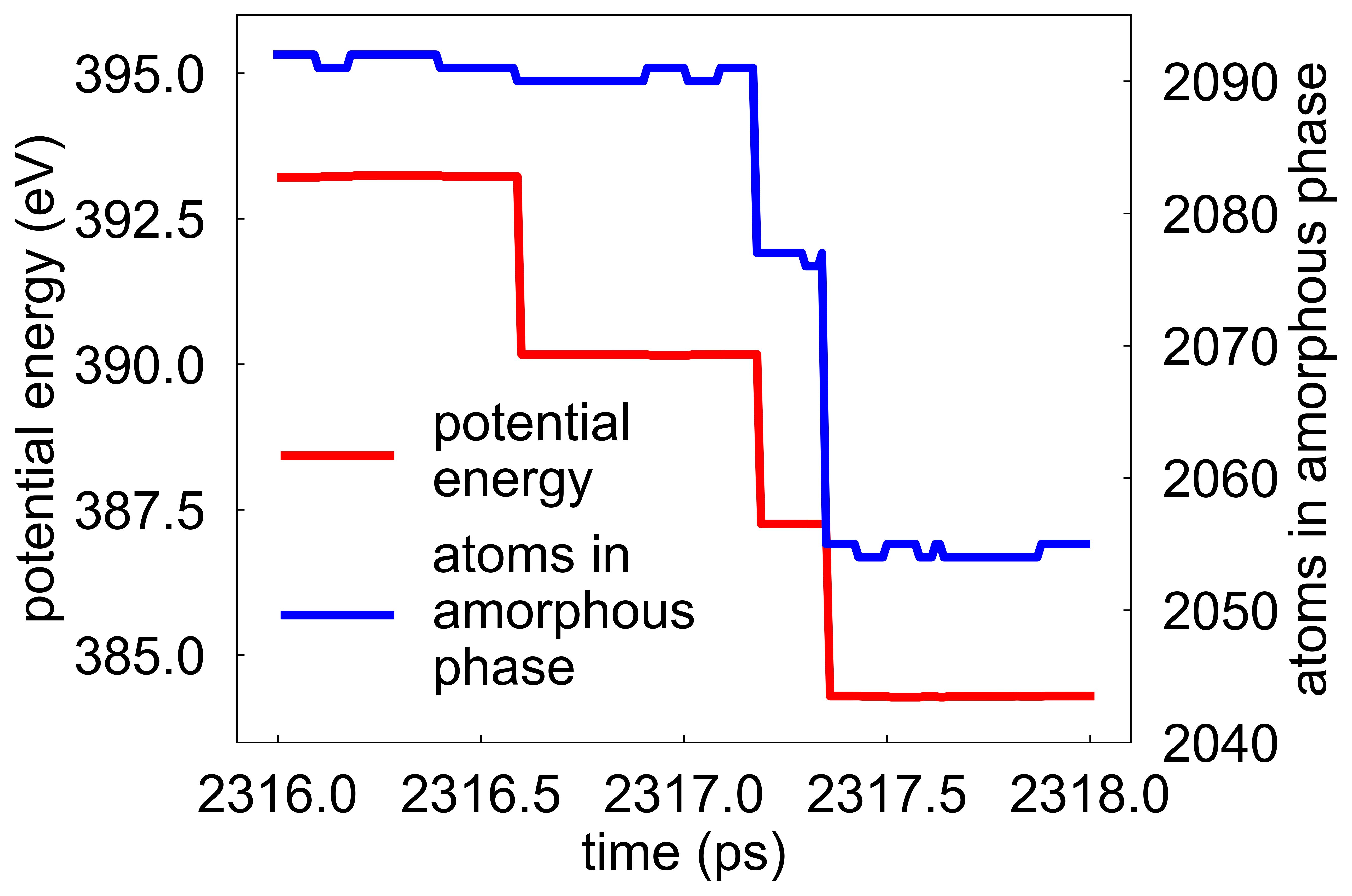} 
&
b) 
\includegraphics[width=0.95\columnwidth]{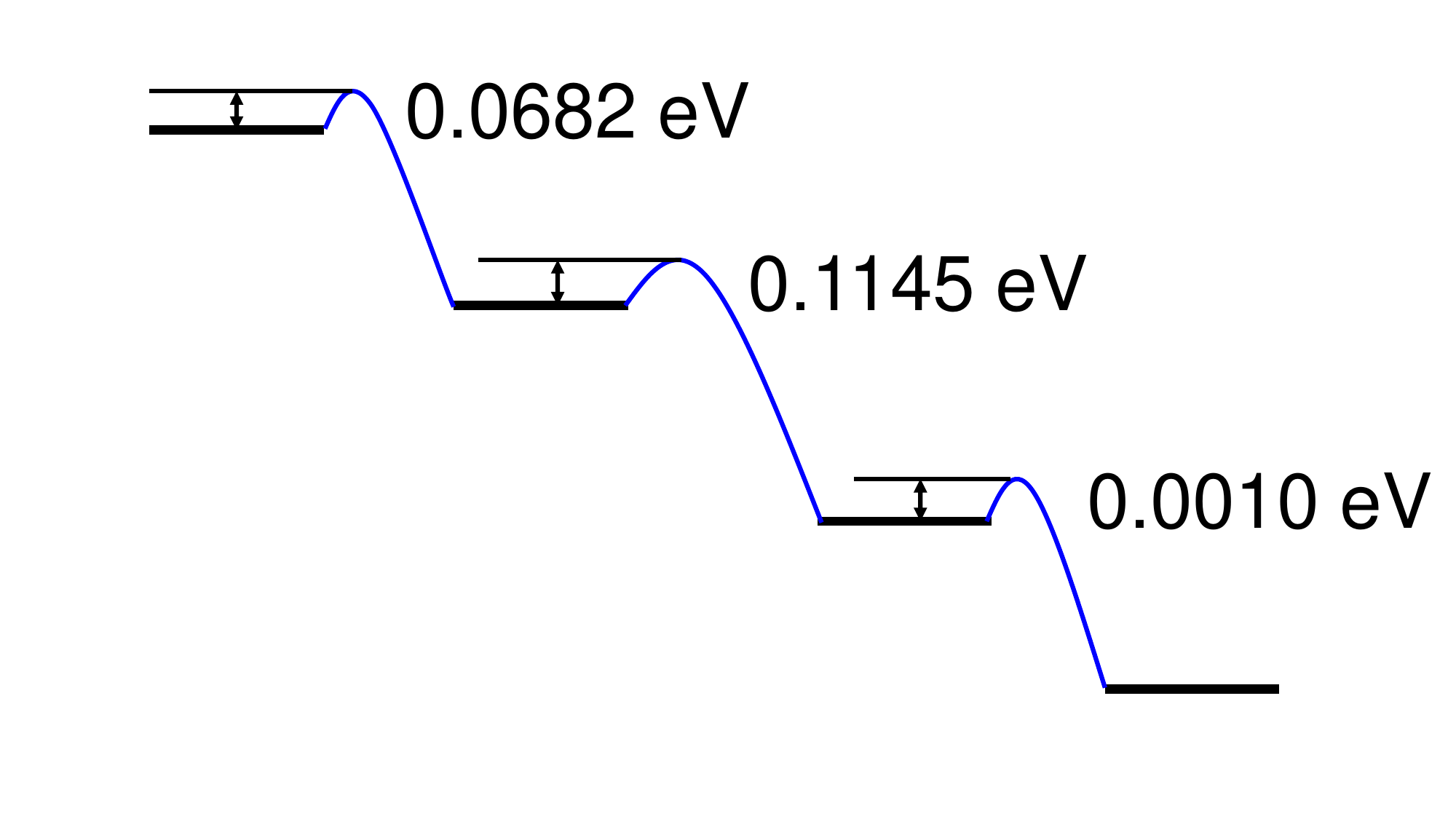} 
\\
c) 
\includegraphics[width=0.95\columnwidth]{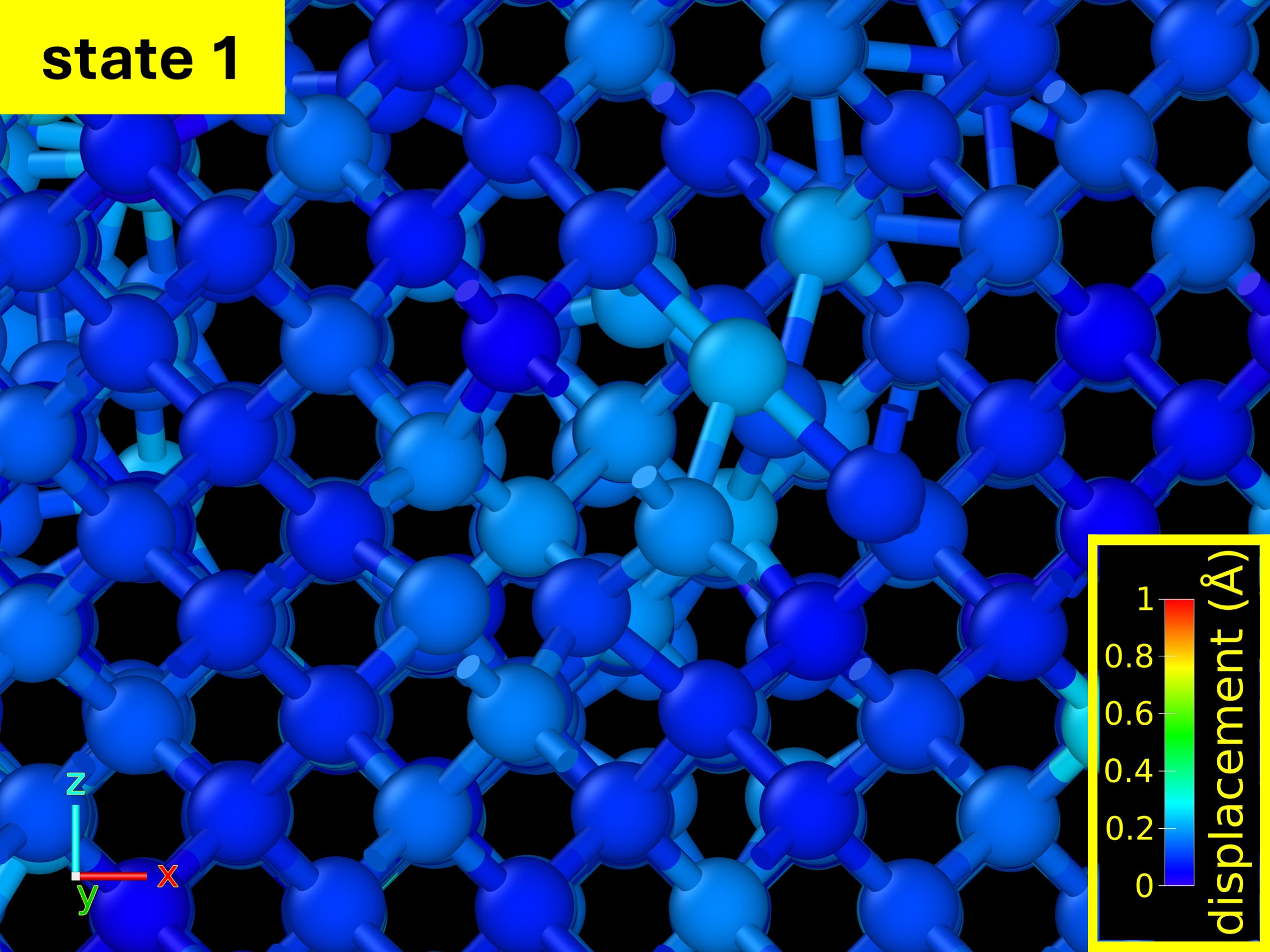} 
&
d)
\includegraphics[width=0.95\columnwidth]{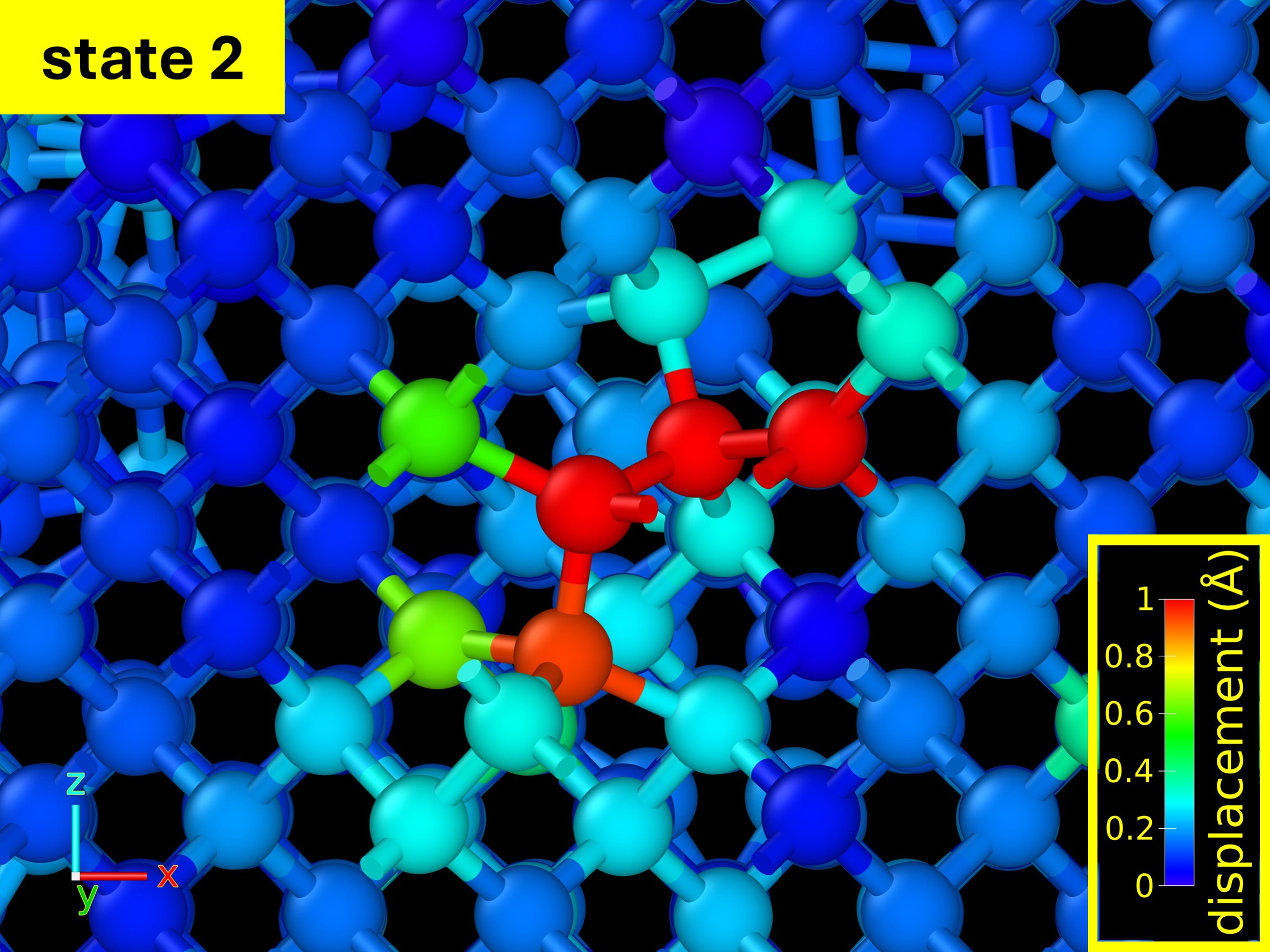} 
\\
e) 
\includegraphics[width=0.95\columnwidth]{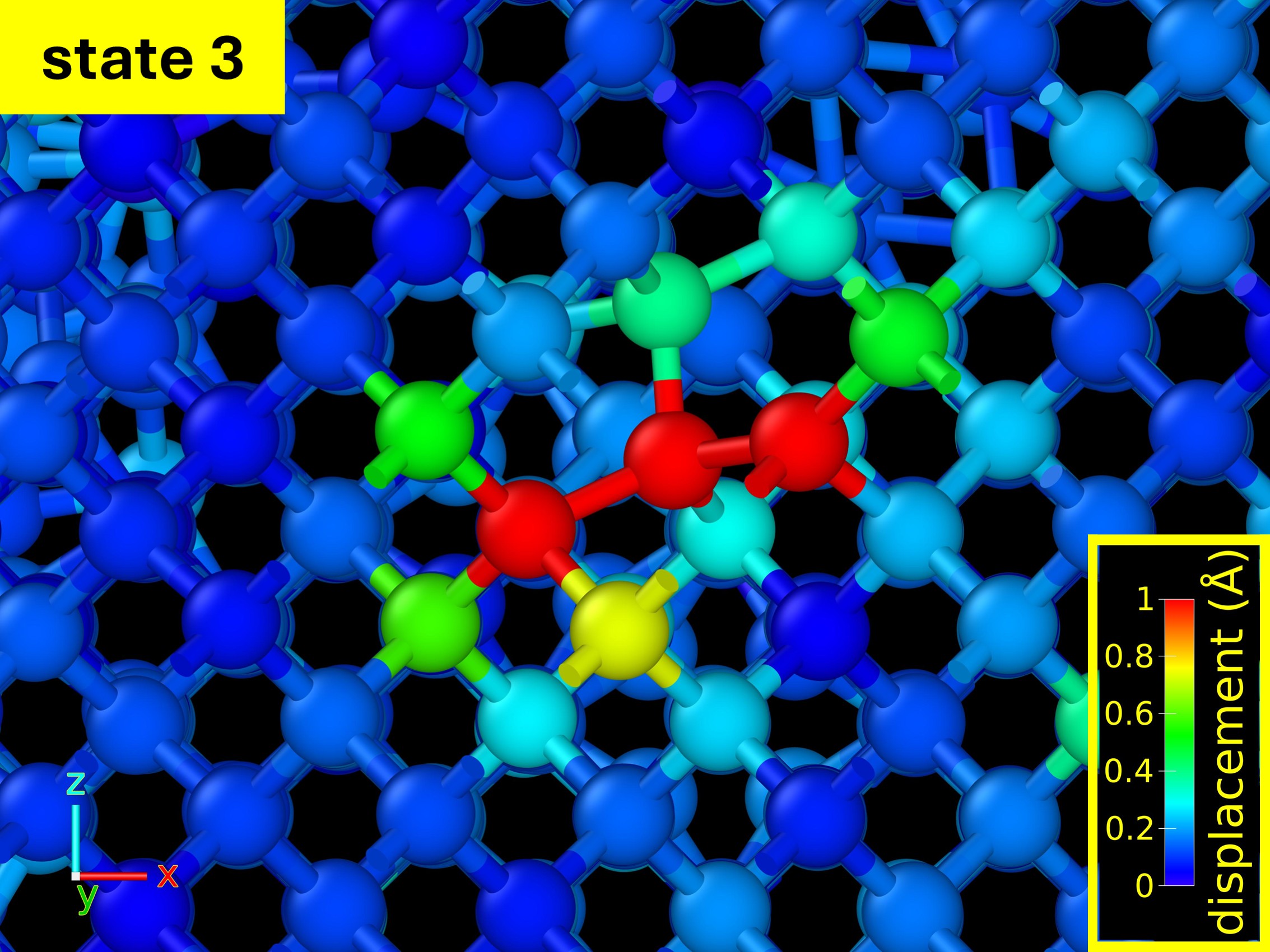} 
&
f)
\includegraphics[width=0.95\columnwidth]{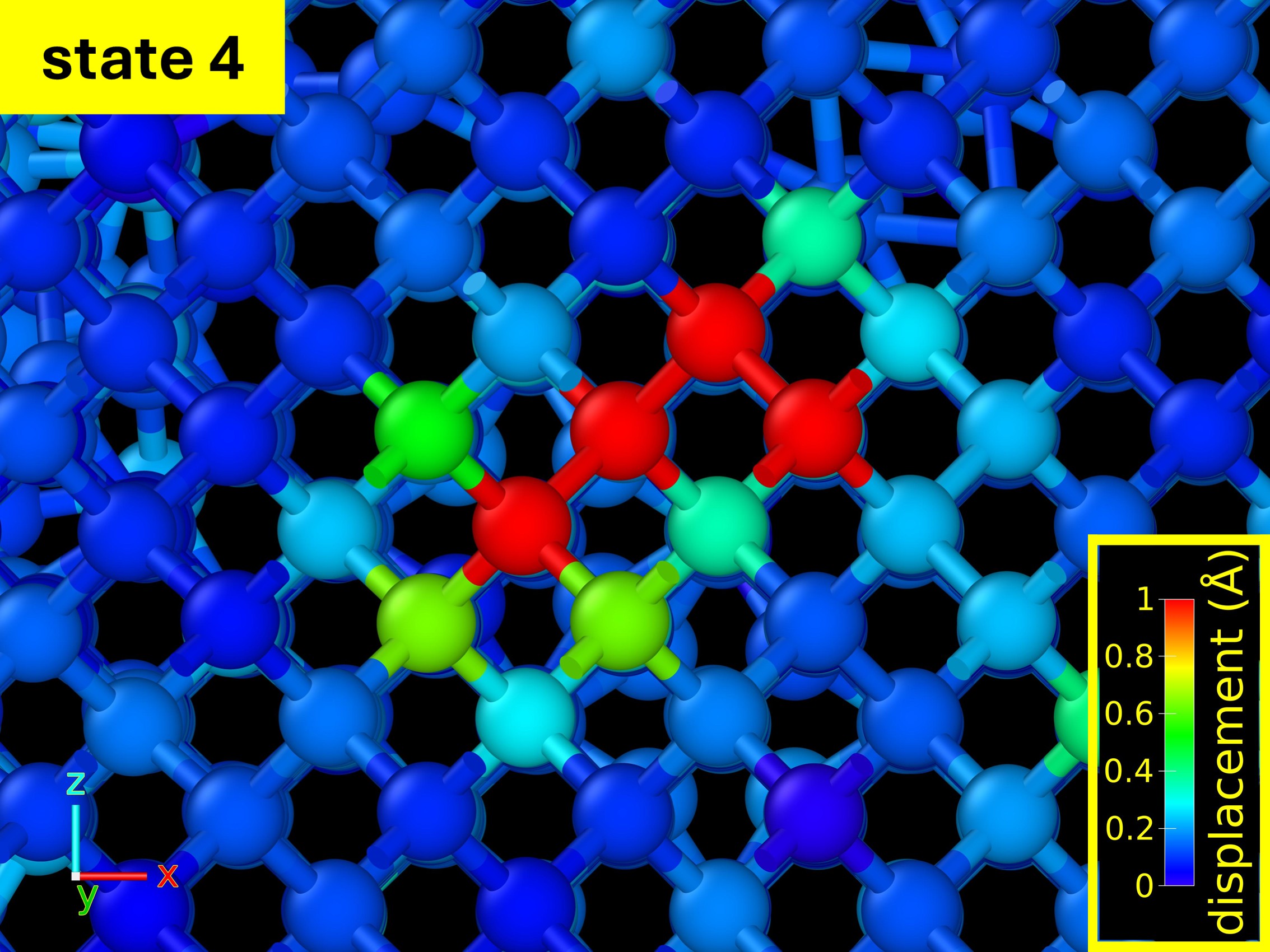} 
\\
\end{tabular}
\end{center}
\caption{\label{fig:annealing_event2}  
An example of the energy barriers in an individual annealing event. In this case the total energy release was 8.92  eV. In (c)(d)(e)(f), the color coding is atomic displacement. 
}
\end{figure*}

Figure \ref{fig:annealing_event1} and Figure \ref{fig:annealing_event2} shows two energy release events whose energy release are larger than 8 eV, taking place at 0.05 K (QTB thermostat).
{\red Figure \ref{fig:annealing_event1} (a) and Figure \ref{fig:annealing_event2} (a) shows the potential energy and the number of atoms in the amorphous phase as functions of time. To get these figures, all atomic configurations have gone through quenching. The number of atoms in the amorphous phase is obtained by DXA analysis in Ovito \cite{ovito}). Figure \ref{fig:annealing_event1} (b) and Figure \ref{fig:annealing_event2} (b) show the energy barriers of the fundamental steps in the two events, which are obtained by climbing image nudged elastic band (CI-NEB) method. Figure \ref{fig:annealing_event1} (c)(d)(e)(f) and Figure \ref{fig:annealing_event2} (c)(d)(e)(f) show the motion of atoms in these two events by color coding of atomic displacement (The zero point of atomic displacement is the atomic configuration before the energy release event). In each figure, the states in (c)(d)(e)(f) take place chronologically, from state 1 in (c) to state 4 in (f). These four states correspond to the four energy minima along the reaction path of the energy release events. }

{\red
From Figure \ref{fig:annealing_event1} (a) and Figure \ref{fig:annealing_event2} (a), one can see that instead of releasing energy all at once, both of the energy release events involve three fundamental steps and four energy minima along the reaction paths. One can observe that the energy releases in the events are often accompanied by the reduction of the number of atoms in the amorphous phase, namely, the crystallization of amorphous pockets. From Figure \ref{fig:annealing_event1} (b) and Figure \ref{fig:annealing_event2} (b) we see that most, if not all the energy barriers of the fundamental steps are lower than 0.1 eV, making it possible for the energy releases to take place at very low temperatures. Moreover, from Figure \ref{fig:annealing_event1} (c)(d)(e)(f) and Figure \ref{fig:annealing_event2} (c)(d)(e)(f), both events involve the migration of several atoms, not just one or two of them. This means that the evolution of defects in these energy release events is not the migration of point defects, but the rearrangement of many atoms. 
}

The results show that at cryogenic temperature, all energy barriers for annealing are very small, and most, if not all of them, are below 0.1 eV. {\red If a metastable configuration overcomes a small energy barrier at first, additional atom rearrangement nearby can be triggered due to the heat released in the first step (overcoming the initial small barrier) in the defect evolution process, and much more energy can be released, which can be viewed as an ``avalanche" effect, just like a terrible avalanche on a snow mountain being triggered by tiny perturbations somewhere on the hill.} In other words, the effect is a variety of self-organized critical phenomena, which are well known e.g to explain dynamics in sandpiles \cite{Bak87,Car90}.

Since the primary radiation damage event will lead to a multitude of complex damage configurations, there is no minimum temperature at which annealing can occur, since there will be barriers of all height present. Therefore, energy release avalanches can be triggered at any temperature. 
{\red Moreover, when compared with the results of B\'eland {\it et al.} and Jay {\it et al.} \cite{Bel13,Bel15,Jay17} showing relaxation events up to second timescales with the k-ART approach, one can conclude that annealing events may occur up to fully macroscopic times.
}

\section{Analysis of location of annealing events}
\label{sec:location_energy_release}

{\red

To analyze where the energy is released in annealing, six annealing cases at 0.05 K were selected, and in total 35 energy release events were collected from these six cases. We visualized all of these events, adding color coding to the atom displacements, to find the atoms that moved the most in energy release events. The reference zero for atom displacements is the atom positions at the beginning of the 2 ps annealing interval where the energy release event takes place. We also used defect meshes from DXA analysis in Ovito \cite{ovito} to mark the boundaries of amorphous pockets in silicon. 

From the visualizations, we discovered that the energy release events took place in the amorphous pockets in damaged silicon. Some of them took place at large amorphous pockets with diameters larger than 10 \text{\AA}, while others took place at quite small amorphous zones. Some of the energy release events took place at the boundaries of the amorphous pockets, which is the amorphous-crystalline interface. The other energy release events took place inside the amorphous pockets. Both of these situations can be viewed as rearrangements of atoms in the amorphous pockets.

A few examples are shown in the following figures. In these figures, atoms are colored by their displacements, and white meshes (defect meshes) mark the boundaries of amorphous pockets. Figure \ref{fig:energy_release_locations} (a) shows an energy release event that takes place at the boundary of a large amorphous pocket, where the proximity of the atom movements to the pocket boundary is clearly shown by the colors of the atoms and the white defect mesh. Figure \ref{fig:energy_release_locations} (b) shows an energy release event that takes place inside a large amorphous pocket. This event can be viewed as a rearrangement of atoms inside an amorphous pocket. Figure \ref{fig:energy_release_locations} (c) shows an energy release event taking place at a small amorphous zone. This small amorphous zone shrinks even further in this energy release event.
}

{\red 
Based on this discovery about the location of annealing events, we can state that the disordered regions in dark matter detector materials, for example, amorphous zones, amorphous-crystalline interface, grain boundaries, interfaces between two different materials, etc., can possibly be active even at cryogenic temperatures. It is likely that defect evolution can take place in these disordered regions, releasing energy that brings disturbance to dark matter direct detection experiments, especially those which are based on phonon detection. This implies that this energy release should be taken into account when building the next generation dark matter detectors. Measures that can be considered to avoid the emergence of atomically disordered regions include optimizing the detector production practice, adopting new manufacturing methods that use less materials with disordered regions, as well as setting up insurmountable barriers between the active detection zone (responsible for interacting with dark matter particles) and other supporting parts of the equipment to block the transmission of the disturbing signals.
}

\section{Estimate of sources and rates of radioactive damage in detectors}

{\red 
As we have shown, the shape of the energy spectrum of annealing events matches well with the excess background observed in the low threshold experiments. In this section we discuss the magnitude of the event rate that could be associated to annealing of radiation damage in silicon detectors. We begin with an estimate of the total power associated with the excess rate, which we obtain by integrating the differential rate shown in Figure 2 of reference \cite{SuperCDMS:2020aus}. This yields an order of magnitude estimate for the excess power per unit detector mass:
\begin{equation}
    P_{\rm excess}\approx 10^5 \frac{\rm eV}{\rm g\, day}.
\end{equation}

This estimate seems compatible also with the recent observations in two different sized silicon detectors in \cite{Chang:2025gkn}. A similar excess rate was also observed in a deep underground measurement with a silicon detector in \cite{Angloher:2022pas}.


If the low energy excess is interpreted as arising from the annealing events, the defect concentration must be replenished in the crystal at approximately the same rate as it is being annealed, otherwise the rate would eventually decay. Therefore, the same power $P_{\rm excess}$ needs to be injected and stored in the crystal structure. 

There are many known sources of trace radioactivity in dark matter detectors. We consider here some of those that have been observed in experiments and given upper limits \cite{DAMIC:2020wkw}.

The most common decay modes of the known radioactivity sources are $\alpha$, $\beta$ and $\gamma$ decay. With energies of the emitted particles typically in the MeV range, conservation of momentum implies that also the nucleus from which the recoil occurs receives kinetic energy and may cause damage. 

For the $\alpha$ decays, for a known emission energy $E_{kin,\alpha}$ the momentum of the emitted particle is (since the MeV kinetic energies are well below the rest mass of the particles $\gg$ 1 GeV, classical kinematics can be used):
\begin{equation}
    E_{\rm{kin},\alpha} = \frac{p_\alpha^2}{2m_\alpha} \Rightarrow p_\alpha = \sqrt{E_{{\rm{kin}},\alpha} 2 m_\alpha}.
\end{equation}

\begin{figure}[H]
\centering
a) 
\includegraphics[width=0.95\columnwidth]{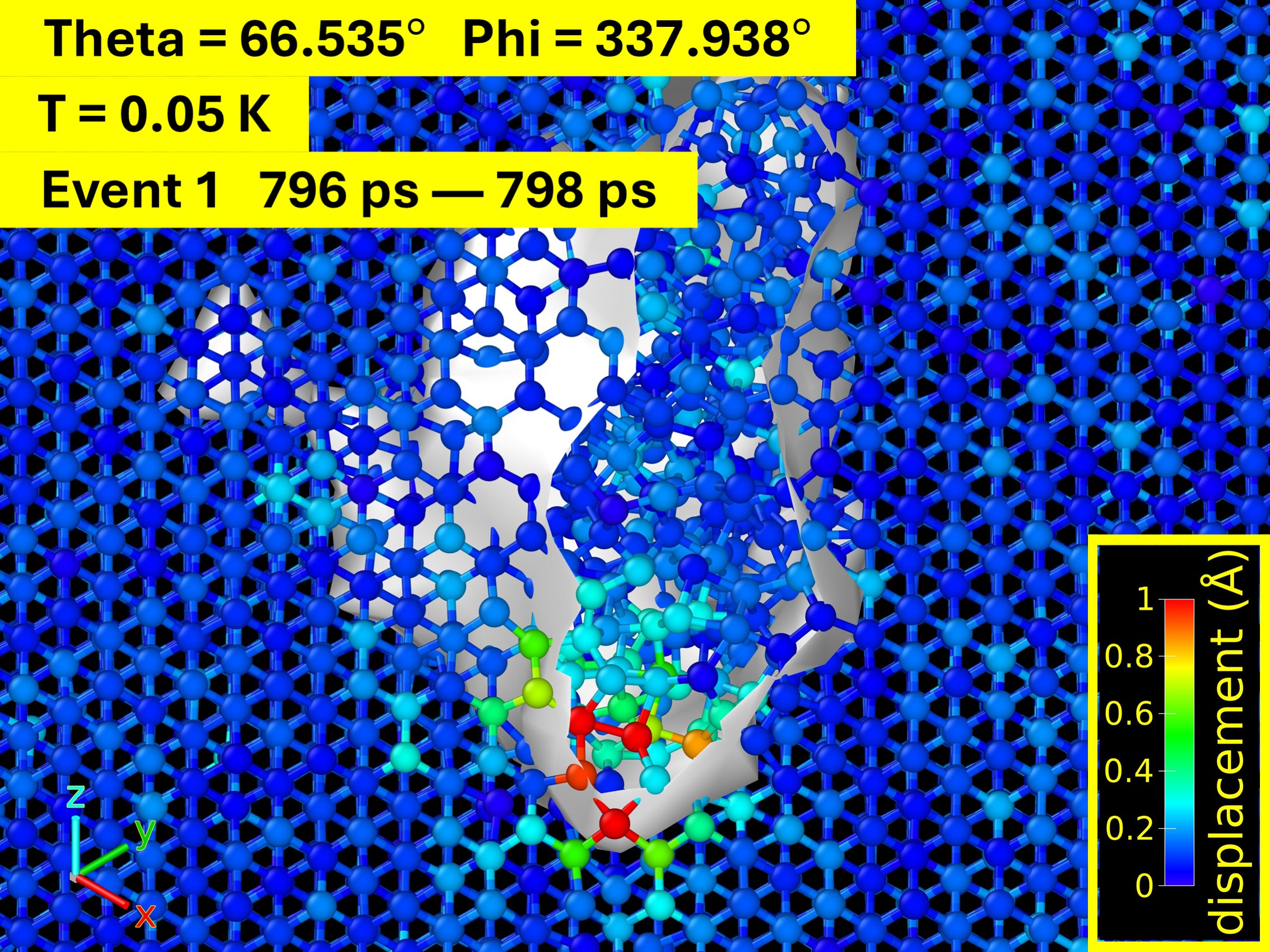} \\
b)
\includegraphics[width=0.95\columnwidth]{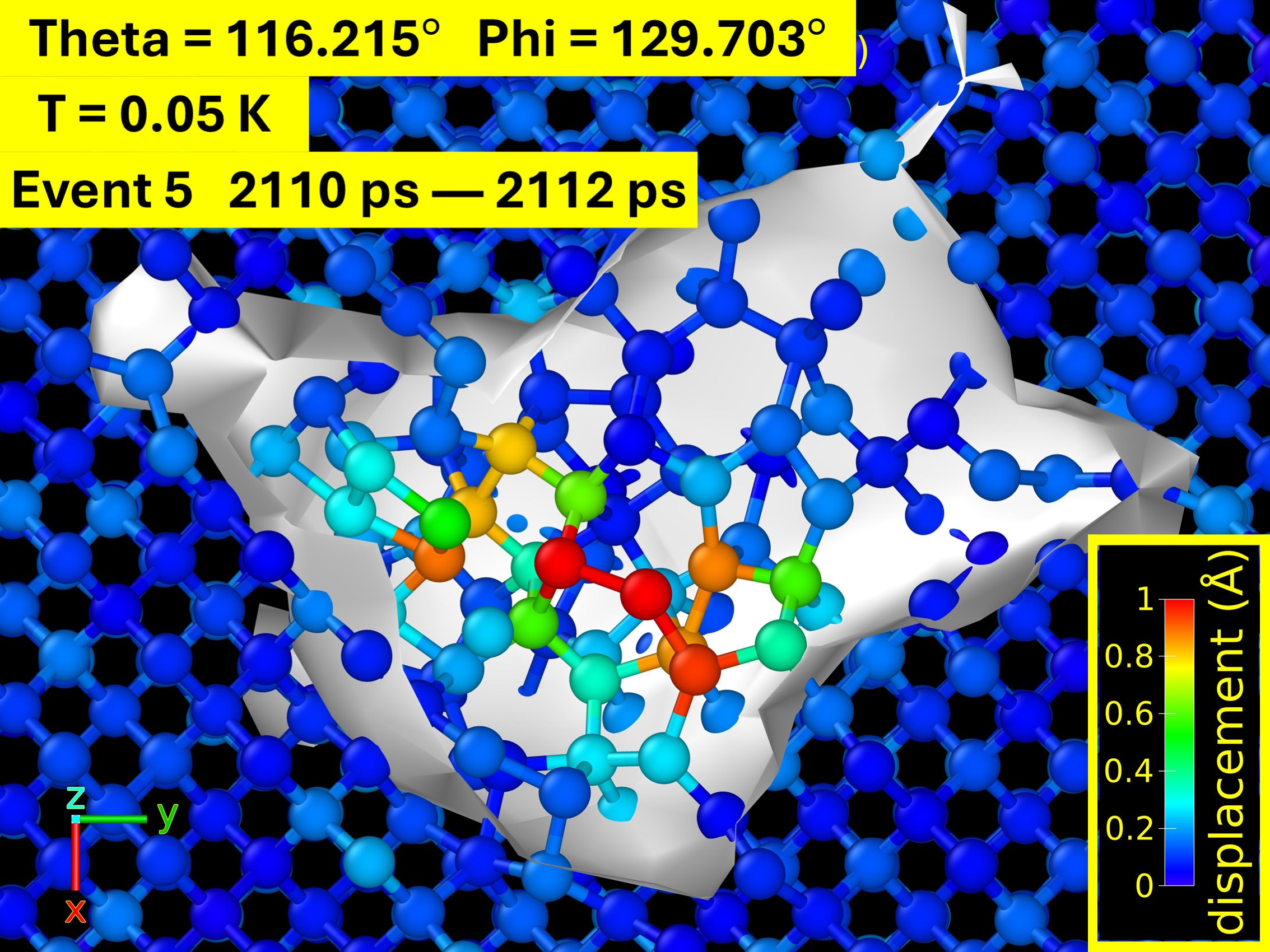} \\
c)
\includegraphics[width=0.95\columnwidth]{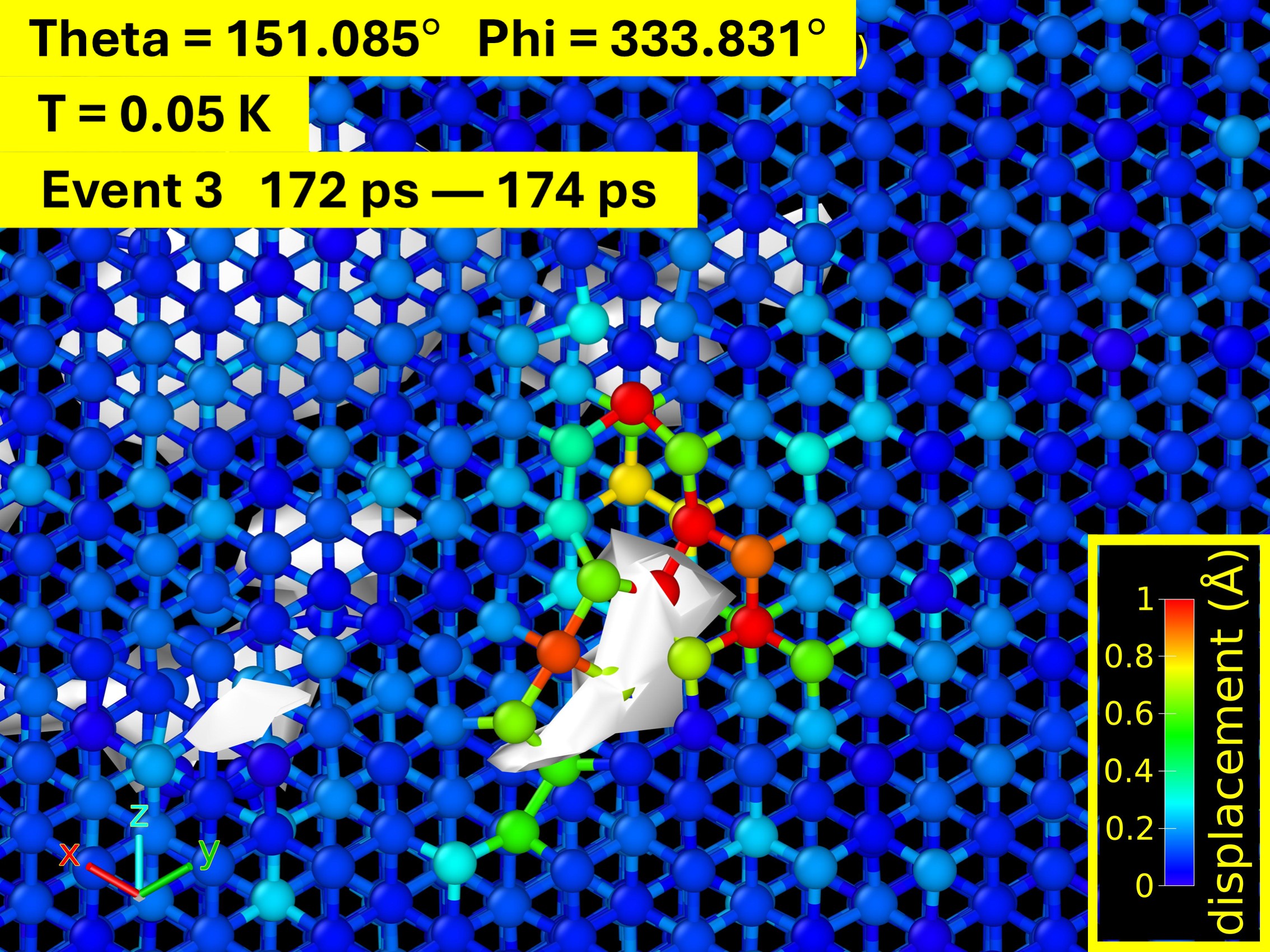} \\
\label{fig:energy_release_locations}
\caption{ 
The locations of three energy release events. The white meshes are defect meshes identified by Ovito DXA analysis, marking the boundaries of amorphous pockets. By adding colors to the atom displacements in the annealing, the place where atoms move a lot (colored in red) can be easily identified, and it is the movement of atoms in these regions that causes the energy release events. Figure \ref{fig:energy_release_locations}(a) is an energy release event taking place at the boundary of an amorphous pocket. Figure \ref{fig:energy_release_locations}(b) is an energy release event taking place inside an amorphous pocket. Figure \ref{fig:energy_release_locations}(c) is an energy release event taking place at a small amorphous zone. Theta ($\theta$) and phi ($\phi$) give the initial recoil direction in each particular cascade simulation shown here.
}
\end{figure}

Here $p_\alpha$ is the momentum and $m_\alpha$ the mass of the $\alpha$ particle.
Due to momentum conservation, the momentum that the decaying nucleus $A$ receives $p_A=p_\alpha$ and hence
\begin{equation}\label{eq:alphaenergytransfer}
    E_{{\rm{kin}},A} =  \frac{p_\alpha^2}{2m_A} = E_{{\rm{kin}},\alpha} \frac{m_\alpha}{m_A}.
\end{equation}

We first consider the decay chain starting from $^{210}{\rm Pb}$ leading to a 5.3 MeV $\alpha$-decay of $^{210}{\rm Po}$. In this process, using eq. \ref{eq:alphaenergytransfer} one obtains
$E_{{\rm{kin}},\rm 206Po}$ = 103 keV \cite{Agu22}. Both the $\alpha$ particle and Pb recoil will in turn lose energy both to electronic and nuclear stopping, out of which only the latter produces significant damage in Si. 

To calculate the nuclear deposited energy $F_{D_n}$, we used the MDRANGE code \cite{Nor94b,MDRANGE} in a non-channeling direction using the NLH interatomic potentials \cite{Nor25} to describe the nuclear collisions. We also checked that the SRIM code \cite{SRIM-2013} gave about the same nuclear deposited energies. The MDRANGE code gave $F_{D_n} \approx 16$ keV for the $\alpha$ and  $F_{D_n} \approx 93$ keV for the Po recoil. Hence the total nuclear deposited energy available for damage production is $F_{D_n} \approx 110$ keV. Comparison with Figure \ref{fig:annealing_differentT} and the data for Si in Ref. \cite{Sas22} shows that about 8\% out of the recoil energy is stored in defects. Thus the  stored energy $E_{\rm stored}\approx 9$ keV.

We now use this value to calculate the expected excess power for the possible upper limit of $^{210}{\rm Po}$ activity
 \mbox{$A = 160 \mu{\rm Bq}/{\rm kg} \approx 0.014/{\rm g\, day}$}
 (\cite{DAMIC:2020wkw}). For the Po decays, the predicted excess energy release would be  
\begin{equation}
    P_{\rm excess, 210Po} = E_{\rm stored} A \approx 130 \frac{\rm eV}{\rm g\, day}.
\end{equation}
In other words, the Po recoils could explain about 0.1\% of the observed excess signal.


We also considered the other parts of the decay chain of $^{238}$U. The chain before $^{210}{\rm Pb}$
has in total seven $\alpha$ decays \cite{DAMIC:2020wkw}, each giving a recoil energy around 100 keV to the parent nucleus. The total recoil energy of these decays is about 670 keV. For the $^{232}$Th chain, the five $\alpha$ particles give in total about 490 keV to parent nucleus recoils.
The 8\% fraction of stored energy $E_{\rm stored U, Th}$ from these recoils is then
$\sim 93$ keV.
The experimental activity limits for these chains are of the order of $A_{U,Th} \approx 0.001/({\rm g\, day})$ \cite{DAMIC:2020wkw,Agu22}, so one can roughly estimate they could contribute 
\begin{equation}
    P_{\rm excess U, Th} = E_{\rm stored U, Th} A_{U,Th} \sim 90 \frac{\rm eV}{\rm g\, day},
\end{equation}
i.e. of the same order as the Po decays, but still not nearly enough to explain the whole signal.



Due to the low mass of the emitted particles, the recoil received from $\beta$ and $\gamma$ decays are small, $\sim$ 100 eV. On the other hand, the 
activities from some of the nuclei decaying by these modes
are rather high. Hence we also calculated roughly the excess yield expected
from one of these. For $^{22}$Na, the dominant decay mode is positron emission 
with a kinetic energy of the positron $\sim 540 $ keV, followed by emission
of a 1274 keV $\gamma$. The recoil energies of these processes to the nuclei
are about 21 and 40 eV, respectively. For these low energies, practically all the energy goes to nuclear deposited energy, and hence the stored energy
can be estimated as (21+40) eV$\times$ 8\%. Using the reported
best-fit activity for \cite{Agu22}  $^{22}$Na of $A_{22Na} = 340 \mu$Bq/kg, one obtains
\begin{equation}
    P_{\rm excess 22Na} = E_{\rm stored 22Na} A_{22Na} \sim 0.14 \frac{\rm eV}{\rm g\, day}.
\end{equation}

A similar calculation for the $\beta$ decays of $^{32}$Si and tritium $^3$H gave similarly small values. 

At sea level, also cosmic muons and neutrons may produce a lot of damage in the samples; however, at the deep underground conditions where the DM detectors are operated, their direct contribution is negligible \cite{Agu22}. They may indirectly contribute by producing radio-active isotopes such as $^{32}$Si when the sample material was on the ground \cite{Sal20}; this contribution is included in the background model \cite{Agu22} that was used to obtain the activities above. We note that a recent observation by the NUCLEUS collaboration \cite{NUCLEUS:excessWS2025} finds some apparent difference on the excess rate between surface and underground measurements, but their analysis on the effect does not support the interpretation that the difference could be attributed to cosmic muons. 

Finally, we note that in the detectors also decay from components close to the actual detector can be a cause of radioactive background. For instance, for the DAMIC detector, detailed background model calculations showed that events in surrounding copper components dominated the background in some energy intervals and could be comparable to the signal in the detector material and surface (Figure 9 in Ref. \cite{Agu22}). As the goal of the current paper is not to analyze a specific detector, we do not aim at detailed calculations of these effects, but just note  that it is reasonable to assume that stored energy caused by radioactivity from the detector surroundings could lead to an excess signal comparable to that from the material itself, i.e. $\sim$ 0.1 - 0.3\% 
of the total.

Taken together, these calculations show that relaxation of energy stored in defects caused by known sources radioactivity in the detectors might explain
excess signal powers $\sim$ 200 - 500 eV/(g day) in Si-based detectors, where the lower limit corresponds to our calculation of known sources of radioactivity at the detector material and surface, and the upper limit comes from adding a very rough estimate of the contributions from surrounding materials.
This level is experimentally significant, but not sufficient to explain the entire signal observed.
Thus it remains to be understood, whether there are other types and sources of damage that could result in similar annealing spectra. For instance, clusters with impurities, inner surfaces of voids, and interfaces between materials are also by definition higher in energy than perfect crystalline bulk, and might be sources of energy release. Systematic consideration of possible energy release
from these is beyond the scope of the current study. However, continuing our work, we aim to consider the material-related effects in more detail in the future.

These calculations were all carried out based on the nuclear energy deposition, since in metals and elemental semiconductors, the electronic energy loss does not produce damage. However, we note that in ionic and amorphous materials also the electronic energy loss can lead to major amounts of damage in the form of swift heavy ion tracks \cite{Tou06,Bie13}, and thus in detectors made of ionic materials, similar levels of radioactivity might cause much higher levels of stored energy that may be released.

}

\section{Discussion and conclusion}
\label{sec:discussion}

The main result of this paper is that {\red annealing} of damage produced by radioactive events can cause energy release on very long time scales after the damage production, even at cryogenic temperatures. The simulations in this work were done on silicon, but considering that it is well known that other covalently bonded semiconductor materials such as Ge and GaAs produce damage in very similar (and even larger) damage pockets as Si \cite{Rua84,Jen95,Jen96,Ben00}, it is very likely that very similar behavior would occur in also other covalently bonded semiconductors.
Many ionic materials also undergo amorphization \cite{Web00,Trach06,Cha13}, and hence a similar mechanisms might be active in at least some of them. Elemental metals cannot be amorphized \cite{Bul56}, however, metastable damage pockets can be formed in them in the form of ordered damage clusters (e.g. C15 structure) and dislocation loops \cite{Sil59b,Bac94,Kir87,Esf21}. However, separate simulation works with appropriate interatomic interaction models for other semiconductors, ionic and metallic materials would be needed to deduce whether energy release from other materials would have a similar energy dependence.

The effect described in this paper has a time dependence on the nanosecond to tens of nanoseconds time scale. Most experiments observing the low-energy excess have not reported such time dependence. This is most likely because their time resolution is large (of the order of milliseconds). Hence if there is sufficient amounts of radioactive impurities or any other damage sources in the sample where numerous decays occur in the detector within the time resolution, the average signal on the long time scales would appear constant, because the detector effectively sums up multiple events.
On the other hand, we do note that the predicted exponential time dependence of the energy release events on time scales of tens of nanoseconds at room temperature might be experimentally testable with time-resolved pulsed ion beam experiments \cite{Wal17,Bar17} and sufficiently fast detectors.

Moreover, as we have shown earlier, a dark matter signal could have a distinct daily variation \cite{Kad17,Hei19,Sas22}, which the excess signals from known sources of radioactive decay would not have.
{\red In case dark matter induces damage pockets in materials similar to those studied in the current work, this part would be expected to have exactly the same energy release behavior as the ones studied here. Hence, while regular matter effects appear to dominate the low energy excess signal, some fraction of it might in principle also have a dark matter origin, and this part would also be expected to have a daily variation.
}

In conclusion, using molecular dynamics simulations of radiation events and their annealing in silicon, we have shown that even at cryogenic temperatures, energy release events of up to 10's of electron volts can be caused by random annealing of complex disordered atom pockets.
We show that the {\red annealing} can occur by avalanche-like events, where a metastable defect configuration first crosses an energy barrier of about 0.1 eV. Crossing this kind of small barrier can then trigger a rapid additional bond {\red rearrangements}, that lead to partial recrystallization and energy releases exceeding 10 eV. This explains why large energy release events can occur even down to cryogenic temperatures. 
These events occur randomly up to macrosopic time scales after the initial picosecond time scale radiation event. Hence, in a sensitive detector they could be caused by annealing of damage produced e.g. by decay of radioactive impurities either in the detector itself or in materials surrounding it. 
These energy release events have an exponential energy dependence that agrees
very well with the experimentally observed low-energy excess energy release in semiconductor detectors.
{\red
Estimation of known sources of  radioactivity in high-quality Si detectors showed that their concentration can explain at most $\sim$ 0.5\% of the observed excess signal, showing that further work is needed to find whether the remaining signal is caused by other structures storing energy, or is due to some completely different effects that also happen to have a similar energy release spectrum.
}

\section*{Ackowledgements}

Grants of computer time from the Finnish IT Center for Science in Espoo, Finland, and the financial support from the Research Council of Finland (grant\# 342777) are gratefully acknowledged.

\newpage

\appendix
\section{Verifying the settings of quantum thermal bath}
\label{appendix:A}

{\red 

There are five parameters to set the quantum thermal bath, which are the target quantum temperature, the temperature damping parameter, the seed for the random number generator to create the random forces, the cutoff frequency of the vibration spectrum, and the number of frequency bins used to discretize the frequency range.

The target quantum temperature is the required temperature for the system, being set at the beginning of annealing simulations. The temperature damping parameter is the reciprocal of the friction coefficient $\gamma$, deciding how rapidly the temperature is relaxed to the target temperature. According to LAMMPS manual \cite{qtb_manual_website}, the suggested range of temperature damping parameter is 0.2 ps to 1 ps, making the friction coefficient $\gamma$ fall into the range of 1 THz to 5 THz. The seed for the random number generator is a positive integer, being used to create the random forces. The cutoff frequency determines the upper limit of the frequency domain of the random forces, and vibrational modes with higher frequencies will not be simulated. According to LAMMPS manual \cite{qtb_manual_website}, the recommended range of cutoff frequency is 2 to 3 times of the Debye frequency of the simulated system. The number of frequency bins sets how many points are sampled in the frequency domain. According to LAMMPS manual \cite{qtb_manual_website}, the number of frequency bins should be large enough so that the classical temperature can converge to the quantum temperature as the temperature increases.

Among these five parameters, the temperature damping parameter and the cutoff frequency have physical significance, so these two parameters should be tested, evaluating how much influence they have on calculation results.

First, the influence of temperature damping is investigated. Figure \ref{fig:heat_capacity_damping_parameter} demonstrates the heat capacity calculation results with different temperature damping parameters. The three heat capacity curves from different temperature damping parameters overlap each other. One can see from this figure that as long as the temperature damping falls in the suggested range given by LAMMPS manual, the influence of temperature damping on the heat capacity results is minimal. Also, all these three heat capacity curves agree with the experimental result (given by Anderson et al. \cite{anderson1930heat}) quite well. 

Figure \ref{fig:temperature_damping_parameter} shows the relationship between the classical temperature and the quantum temperature, demonstrating the influence of temperature damping on it. In this figure, the three temperature curves are very close to each other. One can see that as long as the suggestions from LAMMPS manual is followed, the temperature curve is not sensitive to temperature damping parameter. Therefore, the temperature damping chosen in the simulations with quantum thermal bath in this paper is 0.3 ps, being identical with the temperature damping used in classical thermostats in this paper, enabling a direct comparison between quantum thermal bath results and classical thermostat results.

\begin{figure}[H]
\begin{center}
\includegraphics[width=1.0\columnwidth]{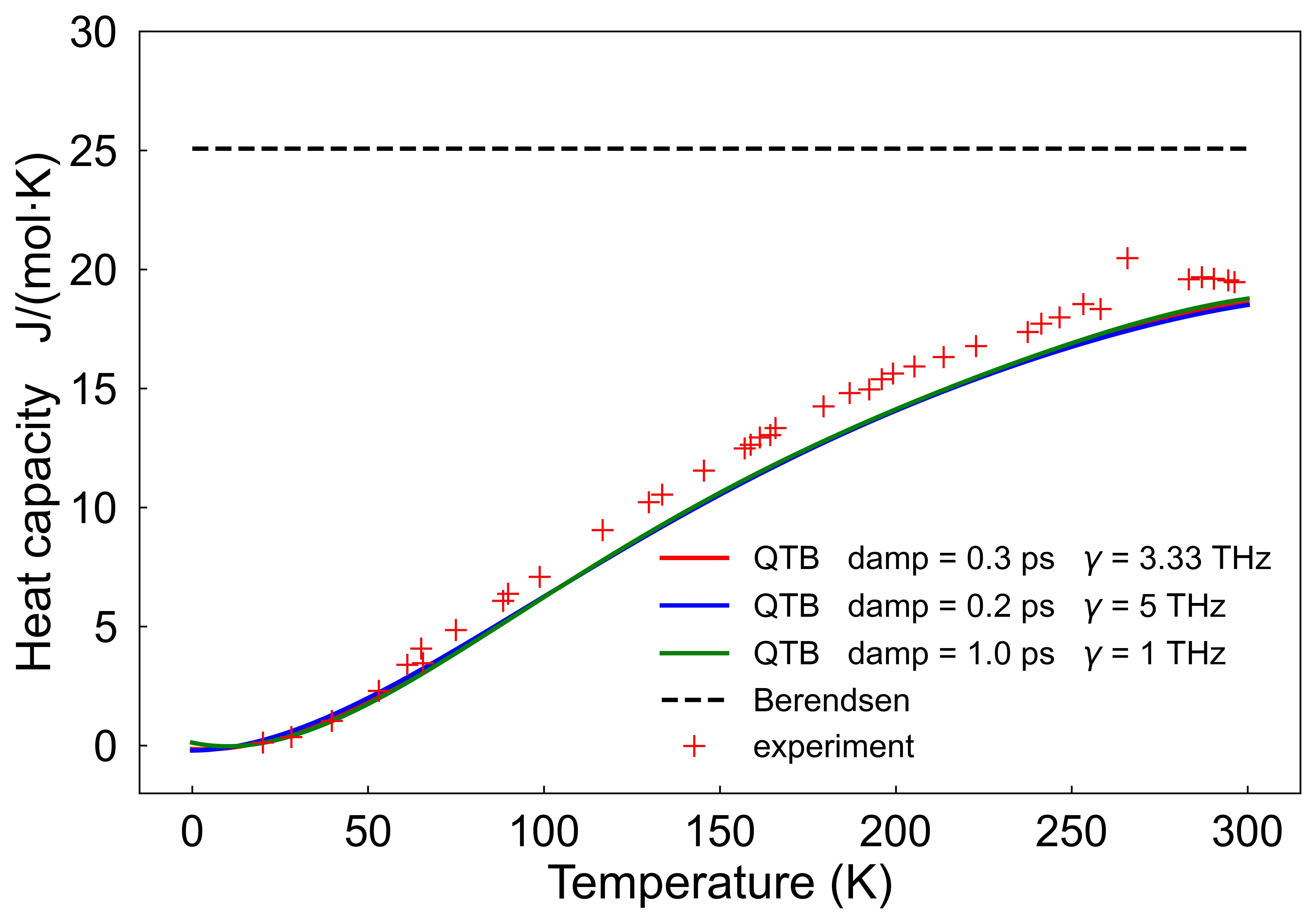}
\end{center}
\caption{\label{fig:heat_capacity_damping_parameter}
\red The influence of temperature damping parameter (``damp" in the figure) on heat capacity results
}
\end{figure}

\begin{figure}[H]
\begin{center}
\includegraphics[width=1.0\columnwidth]{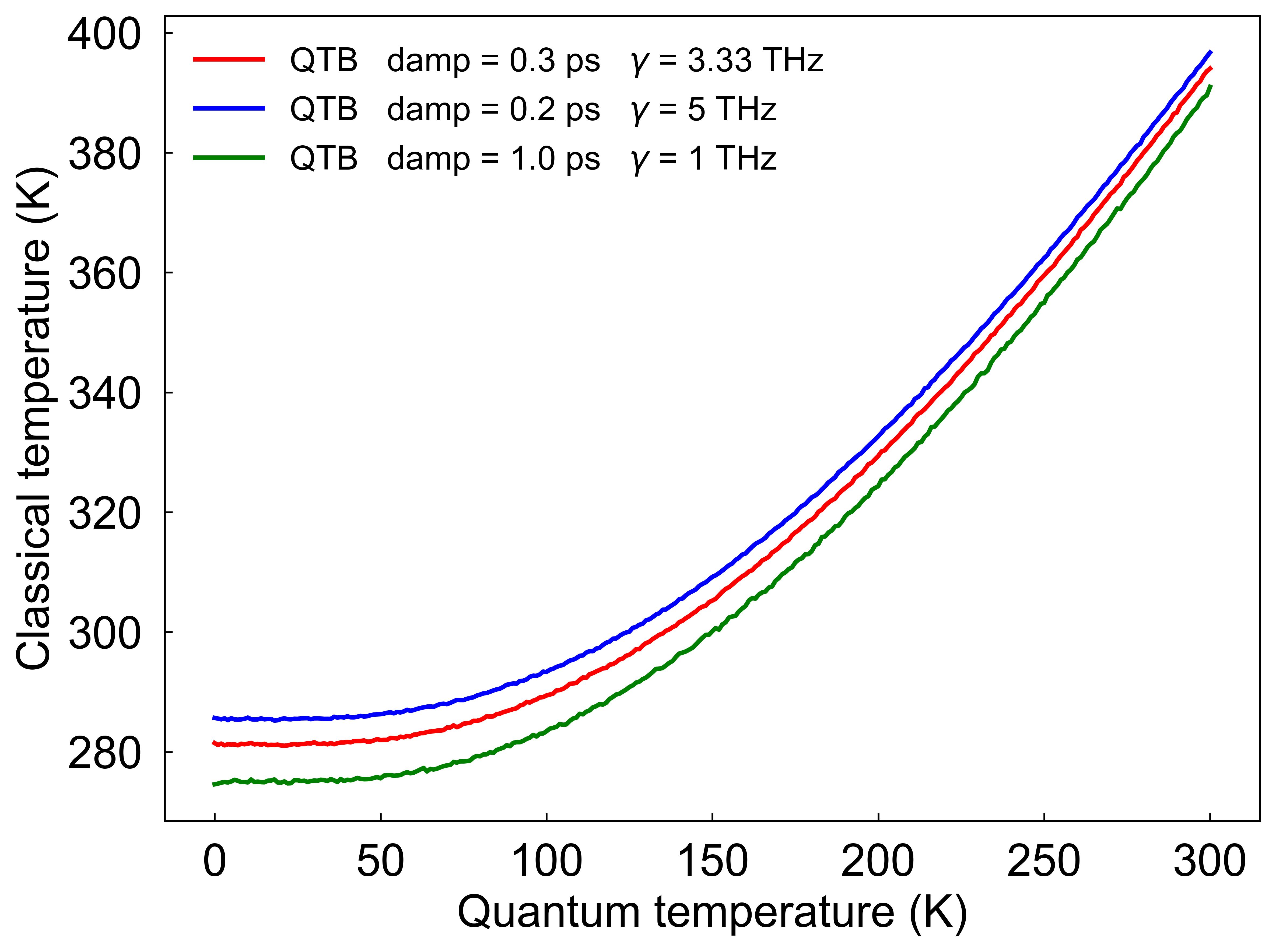}
\end{center}
\caption{\label{fig:temperature_damping_parameter}
\red The influence of temperature damping parameter (``damp" in the figure) on the relationship between quantum temperature and classical temperature
}
\end{figure}

Second, the influence of cutoff frequency is evaluated.
Figure \ref{fig:heat_capacity_cutoff_frequency} demonstrates the heat capacity calculation results with different cutoff frequencies. The three heat capacity curves from different cutoff frequencies overlap each other. This means that as long as the requirements in LAMMPS manual are followed, the heat capacity results are not sensitive to the cutoff frequency. Also,  all of the three curves agree with the experimental result from Anderson et al. \cite{anderson1930heat} quite well.

Figure \ref{fig:temperature_cutoff_frequency} shows the relationship between the classical temperature and the quantum temperature, demonstrating the influence of cutoff frequency on it. In this figure, the three temperature curves from different cutoff frequencies are very close to each other. As long as one follows the suggestions from LAMMPS manual, the temperature curve is not sensitive to cutoff frequency. Therefore, the cutoff frequency chosen in the simulations with quantum thermal bath in this paper is 3 times the Debye frequency of silicon. The reason for choosing this value is that Stillinger-Weber potential overestimates the Debye frequency of silicon by around 17\% \cite{babaei2019machine}. The input cutoff frequency is calculated by three times the Debye frequency coming from the Debye temperature of silicon, which is from reference \cite{silicon_debye_temperature}. This value is smaller than three times the Debye frequency of Stillinger-Weber potential, falling in the suggested range from LAMMPS manual, 2 to 3 times the Debye frequency of the system (Stillinger-Weber potential).

\begin{figure}[H]
\begin{center}
\includegraphics[width=1.0\columnwidth]{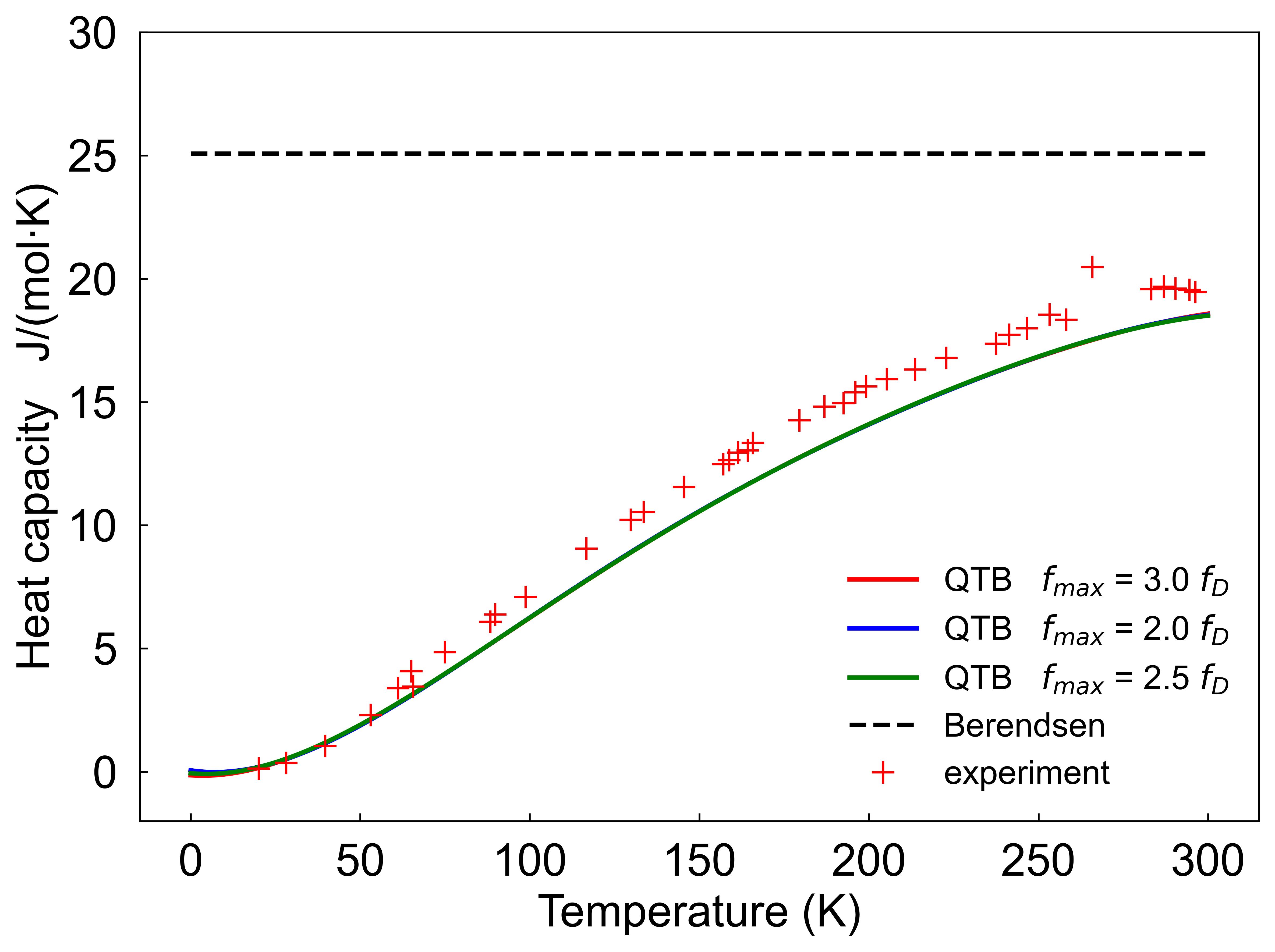}
\end{center}
\caption{\label{fig:heat_capacity_cutoff_frequency}
\red The influence of cutoff frequency ($f_{max}$ in the figure) on heat capacity results
}
\end{figure}

\begin{figure}[H]
\begin{center}
\includegraphics[width=1.0\columnwidth]{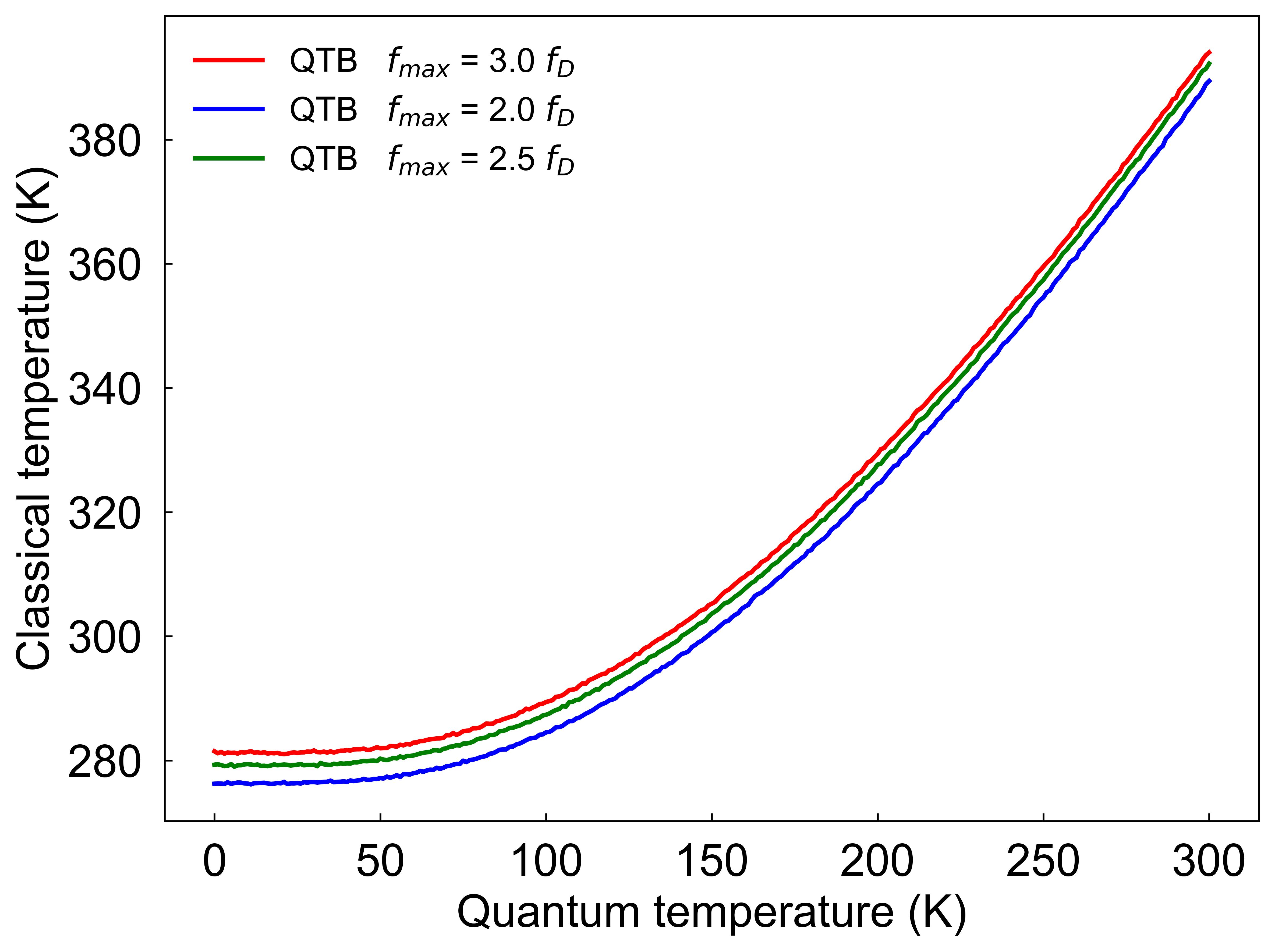}
\end{center}
\caption{\label{fig:temperature_cutoff_frequency}
\red The influence of cutoff frequency ($f_{max}$ in the figure) on the relationship between quantum temperature and classical temperature
}
\end{figure}

Then, with the temperature damping being 0.3 ps and cutoff frequency being 3 times the Debye frequency of silicon, the relationship between heat capacity and temperature is calculated, which is shown in Figure \ref{fig:heat_capacity_800K}. From this figure, the heat capacity curve calculated with quantum thermal bath agrees well with experimental results given by Anderson et al. \cite{anderson1930heat} at low temperatures. Moreover, the heat capacity curve converges to the classical limit given by Dulong-Petit Law at high temperatures. Therefore, one can see that with these settings, the temperature dependence of heat capacity can be correctly reproduced.

\begin{figure}[H]
\begin{center}
\includegraphics[width=1.0\columnwidth]{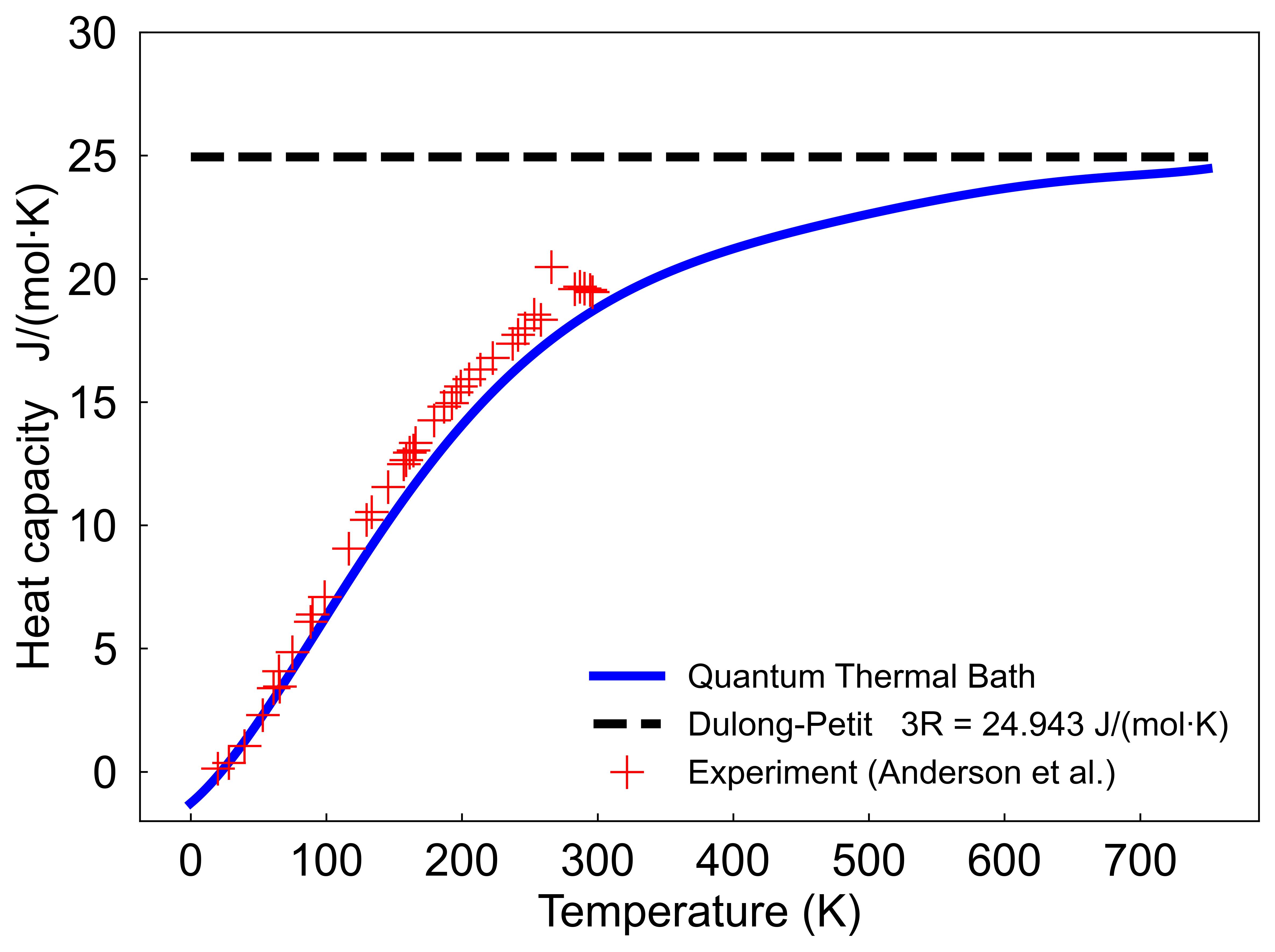}
\end{center}
\caption{\label{fig:heat_capacity_800K}
\red The heat capacity calculation result when the temperature damping parameter is 0.3 ps and the cutoff frequency is 3 times the Debye frequency of silicon.
}
\end{figure}

Finally, the relationship between the classical temperature and the quantum temperature is calculated at a large temperature range from 0 K to 800 K, with the temperature damping being 0.3 ps and cutoff frequency being 3 times the Debye frequency of silicon. The result is shown in Figure \ref{fig:quantum_classical_temperature_800K}. 

\begin{figure}[H]
\begin{center}
\includegraphics[width=1.0\columnwidth]{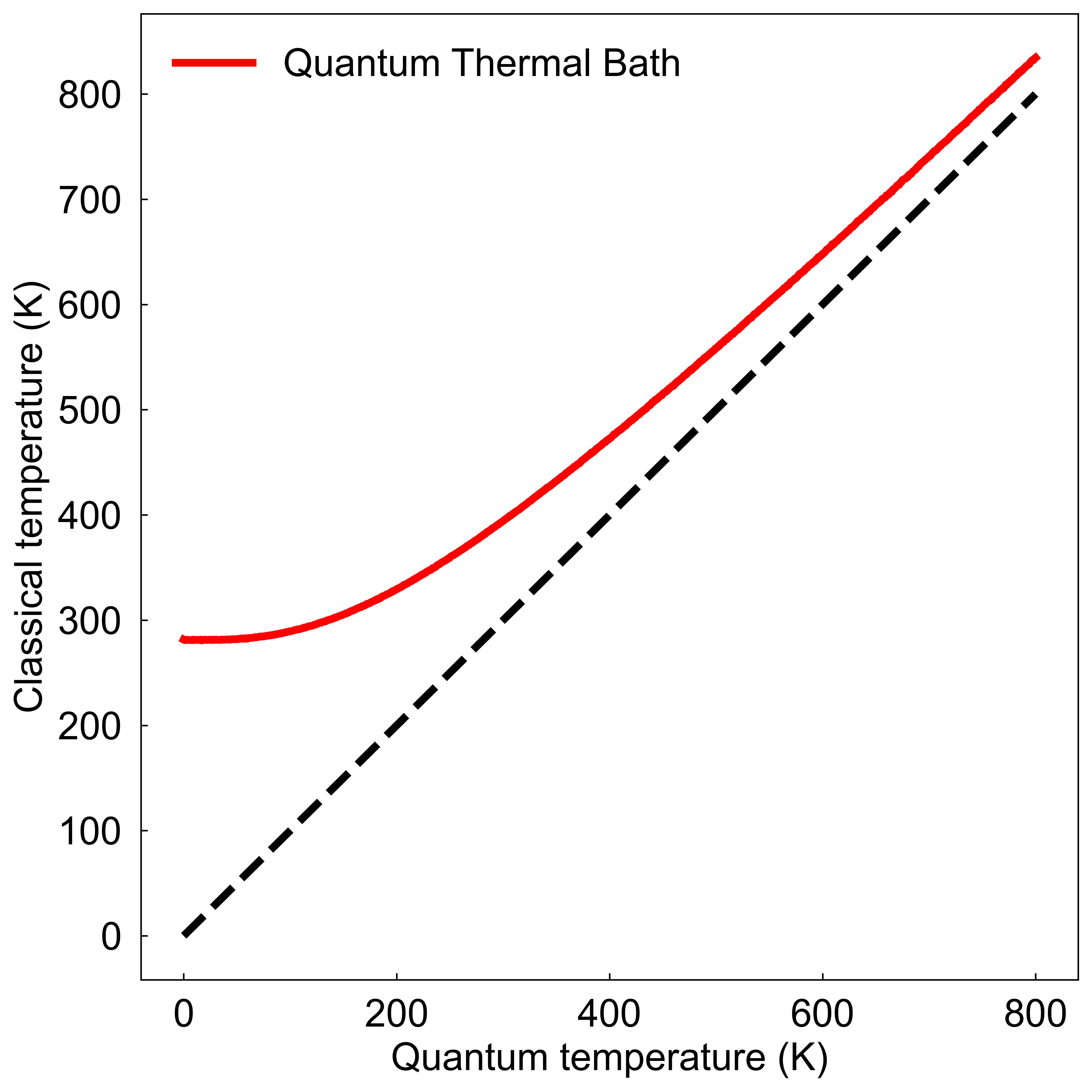}
\end{center}
\caption{\label{fig:quantum_classical_temperature_800K}
The relationship between quantum temperature and classical temperature when the temperature damping parameter is 0.3 ps and the cutoff frequency is 3 times the Debye frequency of silicon. The number of frequency bins is 100.}
\end{figure}

In this figure, a non-zero classical temperature is present even when the quantum temperature approaches zero. This is associated with zero-point energy at low temperatures. At high temperature, the classical temperature gradually converges to the quantum temperature, which is the classical limit. According to LAMMPS manual \cite{qtb_manual_website}, this convergence also justifies the setting of the number of
frequency bins ($N_f$). In the simulations with quantum thermal bath in this paper, the number of frequency bins is set to be 100. This convergence in Figure \ref{fig:quantum_classical_temperature_800K} means that $N_f=100$ is large enough to obtain the convergence of the classical temperature to the target quantum temperature in a high temperature regime.

}

\newpage

\end{document}